\documentclass{article} % For LaTeX2e
\PassOptionsToPackage{table}{xcolor}
\usepackage{amsmath}
\usepackage{titletoc}
\usepackage[utf8]{inputenc}
\usepackage[T1]{fontenc}
\usepackage{graphicx}
\usepackage{booktabs,amsmath, bbm}
\usepackage{multirow}
\usepackage{makecell}
\usepackage{caption}
\usepackage{enumitem}
\usepackage{float}
\usepackage{amsthm}
\usepackage{amssymb}
\usepackage{upquote}
\usepackage{pseudocode}
\usepackage{subfig}
\usepackage{tikz}
\usepackage{algorithm}
\usepackage{algpseudocode}
\usepackage{parskip}
\usepackage{scrextend}
\usepackage{fancyhdr}
\usepackage{pgfplots}
\usepackage{mathtools}
\usepackage[most]{tcolorbox}
\usepackage{xparse}
\usepackage{bm}
\usepackage{pifont}
\usepackage{wrapfig}
\usepackage{placeins}
\definecolor{PennRed}{HTML}{990000}
\definecolor{PennBlue}{HTML}{011F5B}
\definecolor{PennGreen}{HTML}{008e00}
\newcommand{\mycrossmark}{\textcolor{PennRed}{\ding{55}}}
\newcommand{\mycheckmark}{\textcolor{PennGreen}{\ding{51}}}

\usetikzlibrary{arrows.meta, decorations.pathmorphing, fit, backgrounds}

\tikzset{
    myarrow/.style={-{Triangle[length=3mm,width=1mm]}}
}

\definecolor{darkpastelgreen}{rgb}{0.01, 0.75, 0.24}

\newtheorem{innercustomgeneric}{\customgenericname}
\providecommand{\customgenericname}{}
\newcommand{\newcustomtheorem}[2]{%
  \newenvironment{#1}[1]
  {%
   \renewcommand\customgenericname{#2}%
   \renewcommand\theinnercustomgeneric{##1}%
   \innercustomgeneric
  }
  {\endinnercustomgeneric}
}

\newcustomtheorem{ntheorem}{Theorem}
\newcustomtheorem{nlemma}{Lemma}
\newcustomtheorem{nex}{Exercise}

\newtheorem{theorem}{Theorem}

\newtheorem{lemma}[theorem]{Lemma}

\newtheorem{assumption}[theorem]{Assumption}
\newtheorem{approximation}[theorem]{Approximation}

\newcounter{implicitassumption}
\renewcommand{\theimplicitassumption}{A\arabic{implicitassumption}}
\newcommand{\implicitassumption}[2]{%
  \refstepcounter{implicitassumption}\label{#1}%
  \paragraph{(Implicit Assumption \theimplicitassumption) #2.}%
}

\newtheoremstyle{shortassumption}%
  {0.5em}% space above
  {0.5em}% space below
  {\itshape}% body font
  {}% indent
  {\bfseries}% head font
  {.}% punctuation after head
  { }% space after head (single space = inline)
  {}% head spec (empty = default)
\theoremstyle{shortassumption}

\theoremstyle{plain}
\newtheorem*{remark}{Remark}
\newtheorem{proposition}[theorem]{Proposition}

\DeclarePairedDelimiterX{\innerprod}[2]{\langle}{\rangle}{#1, #2}

\newcommand{\norm}[2]{\left\| #1 \right\|_{#2}}
\newcommand{\grad}{\nabla}

\newcommand{\R}{\mathbb{R}}

\newcommand{\KL}[2]{\text{KL}\bigl(#1 \;\|\; #2 \bigr)}

\newcommand{\E}{\mathbb{E}}
\newcommand{\Cov}{\mathrm{Cov}}
\newcommand{\mc}[1]{\mathcal{#1}}

\newcommand{\mbf}[1]{\mathbf{#1}}
\newcommand{\mbx}{\mathbf{x}}
\newcommand{\mby}{\mathbf{y}}
\newcommand{\mbz}{\mathbf{z}}
\newcommand{\mbA}{\mathbf{A}}

\newcommand{\tr}{\text{tr}}

\newcommand{\diag}{\text{diag}}
\newcommand{\eps}{\epsilon}

\newcommand{\ours}{\texttt{ANaLOG}}

\newcommand{\Mconv}{\circledast_{\downarrow M}}
\newcommand{\MconvT}{\circledast_{\uparrow M}}

\usepackage{iclr2027_conference,times}

\usepackage[utf8]{inputenc} % allow utf-8 input
\usepackage[T1]{fontenc}    % use 8-bit T1 fonts
\usepackage{hyperref}       % hyperlinks
\usepackage{url}            % simple URL typesetting
\usepackage{booktabs}       % professional-quality tables
\usepackage{amsfonts}       % blackboard math symbols
\usepackage{nicefrac}       % compact symbols for 1/2, etc.
\usepackage{microtype}      % microtypography
\usepackage{xcolor}         % colors
\usepackage{tcolorbox}

\title{ANaLOG: Anisotropic Native-Latent Operator Guidance for Solving Inverse Problems}

\author{%
    Darshan Thaker, Lachlan Ewen MacDonald \& Ren\'e Vidal \\
    University of Pennsylvania \\
    \texttt{\{dbthaker,lmacdo,vidalr\}@seas.upenn.edu}
}

\iclrfinalcopy % Uncomment for camera-ready version, but NOT for submission.
\begin{document}

\maketitle

\begin{abstract}
Native-latent guidance is a recent paradigm for solving inverse problems with latent diffusion models. It replaces repeated evaluations of the image-space forward model, each requiring a decoder pass, with efficient guidance computed using a learned latent-space surrogate. However, existing methods apply guidance uniformly across latent dimensions, ignoring that measurements are informative only along certain directions and that the reliability of model predictions varies across inputs and timesteps. We propose \ours{}, a framework for efficient uncertainty-aware guidance with pretrained latent diffusion models. \ours{} models uncertainty by learning an anisotropic, input- and time-dependent covariance that is integrated into the guidance mechanism to emphasize reliable directions and downweight uncertain ones. We theoretically analyze this framework in a linear model setting and prove that anisotropic, uncertainty-aware weighting is necessary for correct sampling, whereas isotropic guidance induces sampling errors. Experiments across five challenging inverse problems show that \ours{} improves perceptual reconstruction quality over existing methods while preserving efficiency.
\end{abstract} 
\section{Introduction}
\label{sec:intro}
Many image processing tasks, such as image deblurring and inpainting, can be
formulated as inverse problems. The goal of these tasks is to recover a clean image
$\mbx_0$ from a degraded measurement $\mby \approx \mc{A}(\mbx_0)$ produced by a known forward
operator $\mc{A}$, such as blurring or masking. A popular Bayesian 
approach is to sample from the posterior $p(\mbx_0 \mid \mby)$ using
a data prior $p(\mbx_0)$ and a forward model $p(\mby \mid \mbx_0)$. Pretrained diffusion priors have become the dominant approach 
for modeling $p(\mbx_0)$, where this Bayesian approach becomes akin to
diffusion guidance methods~\citep{chung2022diffusion, daras2024survey}. 

Despite the successes of these methods, there remain unique challenges in developing accurate and efficient diffusion-based inverse problem solvers. One difficulty is computational: visual data are high-dimensional, motivating scalable priors and efficient guidance algorithms. Latent diffusion models (LDMs) provide one such scalable prior by performing diffusion in a compressed latent space rather than in pixel space \citep{rombach2022high}. Another difficulty is statistical: guidance should account for how much information the measurement provides at each timestep, and how uncertain the diffusion model is about the corresponding clean image features \citep{peng2024improving, rissanen2024free}. For example, guidance at a given timestep should emphasize components that are both important to restore under the given forward operator and can be predicted reliably from the current noisy state. Capturing this timestep-, input-, and task-dependent uncertainty is therefore central to reliable guidance.

An ideal method should combine (1) a scalable diffusion prior, (2) efficient guidance, and (3) uncertainty-aware guidance. However, prior work addresses only subsets of these desiderata (Table~\ref{tab:teaser}), and no existing method achieves all three simultaneously. For instance, there is a mature line of work in developing uncertainty-aware guidance algorithms for pixel-space diffusion methods by modeling quantities such as $\Cov(\mby \mid \mbx_t)$ at timestep $t$ in the reverse diffusion process, exploiting linearity of the forward operator and approximations to the posterior covariance of $p(\mbx_0 \mid \mbx_t)$. However, these methods do not extend to LDM priors as the analogous uncertainty is $\Cov(\mby \mid \mbz_t)$, and nonlinearity of the LDM decoder prevents any closed-form computation of this covariance. On the other hand, existing LDM inverse problem solvers fall into two categories: decode-and-guide methods and native-latent methods. Decode-and-guide methods map latents back to image space at each sampling step and apply guidance using the original forward model. This preserves the image-space forward operator, but sacrifices sampling efficiency through repeated decoder evaluations. Uncertainty-aware variants can further increase cost by requiring additional denoiser evaluations at each step~\citep{rout2024beyond}. In contrast, native-latent methods improve sampling efficiency by applying guidance directly in latent space, for example by learning a latent forward operator $H_\theta$ that maps the denoiser-predicted clean latent $\E[\mbz_0 \mid \mbz_t]$ to $\mc{E}(\mby)$~\citep{raphaeli2025silo}. However, native-latent guidance was not derived under a probabilistic framework, leaving open the question of uncertainty estimation and its integration into guidance.

\begin{figure}[t]
    \centering
    \includegraphics[width=\linewidth]{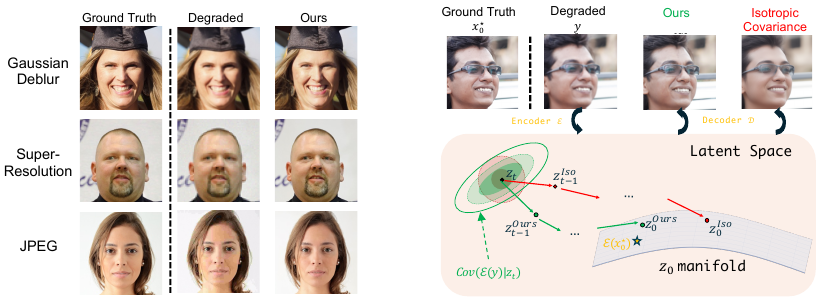}
    \caption{Left: Representative results of our method on the Gaussian deblurring, 8x super-resolution, and nonlinear JPEG decompression tasks. Right: \ours{} models anisotropic covariance during guidance without expensive decoding steps. On a Gaussian deblurring task, \ours{} preserves identity-relevant detail that isotropic approximations wash out.}
    \label{fig:teaser_qual}
    \vspace{-4.5mm}
\end{figure}

\begin{wraptable}{r}{0.45\linewidth}
    \centering
    \vspace{-1mm}
    \scriptsize
    \setlength{\tabcolsep}{3pt}
    \captionof{table}{Comparison of posterior sampling methods across three desiderata.}
    \label{tab:teaser}
    \begin{tabular}{lccc}
        \toprule
        & \makecell{\textbf{Efficient} \\
          \textbf{Diffusion}}
        & \makecell{\textbf{Efficient} \\
          \textbf{Guidance}}
        & \makecell{\textbf{Uncertainty-} \\
          \textbf{Aware}} \\
        \midrule
        Pixel-space
        & \mycrossmark & \mycheckmark
        & \mycheckmark \\
        LDPS, PSLD
        & \mycheckmark & \mycrossmark
        & \mycrossmark \\
        STSL
        & \mycheckmark & \mycrossmark
        & \mycheckmark \\
        SILO
        & \mycheckmark & \mycheckmark
        & \mycrossmark \\
        \textbf{\ours{} (Ours)}
        & \mycheckmark & \mycheckmark
        & \mycheckmark \\
        \bottomrule
    \end{tabular}
    \vspace{-3mm}
\end{wraptable}
In this paper, we bridge this gap and 
introduce \ours{} (\textbf{An}isotropic
\textbf{Na}tive-\textbf{L}atent \textbf{O}perator
\textbf{G}uidance), a framework that allows
for efficient and uncertainty-aware guidance
in the latent space. 
To develop our framework, we first reformulate existing work on native-latent guidance \citep{raphaeli2025silo}
as approximate inference under a Gaussian latent measurement
likelihood with a fixed covariance. We then extend 
to learn richer covariance structure without relying on
covariance propagation through the LDM decoder.
Specifically, our main contributions are: 

\begin{enumerate}[leftmargin=*]
\item \textbf{Probabilistic Framework.} 
%We develop \ours{} as a probabilistic framework for modelling the latent measurement likelihood $p(\mc{E}(\mby) \mid \mbz_t)$
We develop \ours{}, a probabilistic framework in which the latent measurement likelihood $p(\mc{E}(\mby) \mid \mbz_t)$ is modeled
as a Gaussian with learned mean and covariance. Viewed through this lens, existing work \citep{raphaeli2025silo} corresponds to the special case where the covariance is fixed to an isotropic, input-, and time-independent matrix. In contrast, \ours{} learns an anisotropic, input-, and time-dependent covariance via a simple training procedure that augments the learning of the latent forward operator with a lightweight uncertainty head. At inference-time, we develop a guidance algorithm that steers using a Mahalanobis-weighted residual.   

\item \textbf{Theoretical Results.} We theoretically isolate the role of covariance modeling in a tractable setting with Fourier-diagonal data and a linear LDM. We prove, via a lower bound, that even under perfect mean estimation, any isotropic covariance approximation incurs a strictly positive error relative to the true likelihood. In contrast, \ours{} recovers the true likelihood at the standard parametric convergence rate.

\item \textbf{Empirical Evidence.} We empirically show that \ours{} improves perceptual quality over existing LDM guidance algorithms across five inverse problems while remaining substantially faster than decode-and-guide methods. Furthermore, we show that visually, our learned uncertainty estimates closely match important input-, task-, and time-dependent structure.
%input-dependent, task-dependent, and time-dependent structure.
\end{enumerate}

\section{Preliminaries and Related Work}
\label{sec:related_work}

\paragraph{Score-Based Generative Models.} 
Diffusion models~\citep{ho2020denoising, song2020score} are powerful score-based generative models. They define a forward process that gradually corrupts data $\mbx_0 \sim p(\mbx_0)$ into noise, and train a neural network to reverse this process by learning the time-dependent score function $\nabla_{\mbx_t} \log p(\mbx_t)$ \citep{efron2011tweedie}. Sampling proceeds by solving the reverse-time SDE~\citep{anderson1982reverse}:
\begin{equation}\label{eq:reverse_sde}
    \mathrm{d}\mbx_t = \bigl[f(t)\mbx_t - g(t)^2 \nabla_{\mbx_t} \log p(\mbx_t)\bigr]\,\mathrm{d}t + g(t)\,\mathrm{d}\bar{\mbf{w}},
\end{equation}
where $f(t)$ and $g(t)$ are the drift and diffusion coefficients of the forward process and $\bar{\mbf{w}}$ is a standard reverse-time Wiener process. Latent diffusion models (LDMs)~\citep{rombach2022high} apply this framework in the compressed latent space of a pretrained encoder $\mc{E}: \mc{X} \to \mc{Z}$, with a corresponding decoder $\mc{D}: \mc{Z} \to \mc{X}$. Here, we denote $\mc{X}$ as pixel space and $\mc{Z}$ as latent space (typically of dimension smaller than ambient image dimension). The forward process corrupts latents $\mbz_0 = \mc{E}(\mbx_0)$, and the score network is trained to approximate $\nabla_{\mbz_t} \log p(\mbz_t)$.

\paragraph{Posterior Sampling.} For inverse problems, we observe measurements $\mby = \mc{A}(\mbx_0) + \sigma_y \bm{\epsilon}$, where $\mc{A}$ denotes the forward operator and $\bm{\epsilon} \sim \mc{N}(\bm{0}, \mbf{I})$. We first describe the methodology for using pixel-space diffusion priors to solve inverse problems and then highlight key challenges in extending these approaches to LDMs. To solve inverse problems, we wish to sample from $p(\mbx_0 \mid \mby)$, which requires running the reverse SDE~\eqref{eq:reverse_sde} with the \emph{conditional} score. This decomposes via Bayes' rule as:
\begin{equation}\label{eq:latent_bayes}
    \nabla_{\mbx_t} \log p(\mbx_t \mid \mby)
    = \underbrace{\nabla_{\mbx_t} \log p(\mbx_t)}_{\text{unconditional score}}
    + \underbrace{\nabla_{\mbx_t} \log p(\mby \mid \mbx_t)}_{\text{noisy likelihood score}}.
\end{equation}
The unconditional score is provided by the pretrained diffusion prior. Conversely, the noisy likelihood factorizes as $\int p(\mby \mid \mbx_0) p(\mbx_0 \mid \mbx_t) \ d \mbx_0$, which is intractable to analytically compute in general due to the unknown 
posterior denoising distribution \citep{daras2024survey}. This has motivated the development of numerous approaches which focus on approximating $p(\mbx_0 \mid \mbx_t)$, e.g., as a Gaussian distribution with covariance approximated as a diagonal \citep{peng2024improving} or low-rank matrix \citep{meng2021estimating} (see Appendix~\ref{app:related_work} for a more detailed history). Crucially, these approaches rely on linearity of the forward operator in order to compute a closed form of the integral above. 

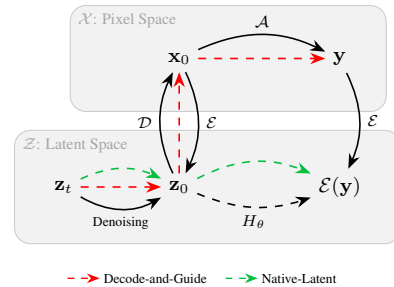
\begin{wrapfigure}{r}{0.37\linewidth}
\vspace{-12pt}
\centering
\begin{tikzpicture}[
    every node/.style={font=\small},
    arr/.style={-{Stealth[length=2mm]}, semithick},
    scale=0.85, transform shape
]
\node (x) at (0,0) {$\mbx_0$};
\node (Ax) at (2.5,0) {$\mby$};
\node (z) at (0,-2) {$\mbz_0$};
\node (Hz) at (2.5,-2) {$\mc{E}(\mby)$};
\node (zt) at (-1.8,-2) {$\mbz_t$};
\coordinate (xleft) at (-1.2,0);

\begin{scope}[on background layer]
    \node[fill=gray!10, rounded corners=6pt, draw=gray!40,
          fit=(xleft)(Ax), inner xsep=12pt, inner ysep=14.5pt] (pixelbox) {};
    \node[anchor=north west, font=\scriptsize, text=gray!70] at (pixelbox.north west) {$\mc{X}$: Pixel Space};
    \node[fill=gray!10, rounded corners=6pt, draw=gray!40,
          fit=(zt)(z)(Hz), inner xsep=12pt, inner ysep=14.5pt] (latentbox) {};
    \node[anchor=north west, font=\scriptsize, text=gray!70] at (latentbox.north west) {$\mc{Z}$: Latent Space};
\end{scope}

\draw[arr] (x) to[bend left=25] node[above, font=\scriptsize] {$\mc{A}$} (Ax);
\draw[arr] (x) to[bend left=25] node[right, font=\scriptsize] {$\mc{E}$} (z);
\draw[arr] (Ax) to[bend left=25] node[right, font=\scriptsize] {$\mc{E}$} (Hz);
\draw[arr, dashed] (z) to[bend right=20] node[below, font=\scriptsize] {$H_\theta$} (Hz);
\draw[arr] (z) to[bend left=25] node[left, font=\scriptsize] {$\mc{D}$} (x);
\draw[arr] (zt) to[bend right=30] node[below, font=\tiny] {Denoising} (z);

\draw[arr, dashed, red] (zt) -- (z);
\draw[arr, dashed, red] (z) -- (x);
\draw[arr, dashed, red] (x) -- (Ax);

\draw[arr, dashed, darkpastelgreen] (zt) to[bend left=25] (z);
\draw[arr, dashed, darkpastelgreen] (z) to[bend left=25] (Hz);

\node[anchor=north, font=\tiny] at (0.35,-3.2) {
    \begin{tabular}{@{}l@{\hspace{8pt}}l@{}}
    \tikz{\draw[arr, dashed, red] (0,0) -- (0.5,0);} Decode-and-Guide &
    \tikz{\draw[arr, dashed, darkpastelgreen] (0,0) -- (0.5,0);} Native-Latent
    \end{tabular}
};
\end{tikzpicture}
\caption{Guidance pathways at each sampling step.}
\label{fig:visual_descrip}
\vspace{-10pt}
\end{wrapfigure}

For LDMs, approximating the corresponding posterior covariance $p(\mbz_0 \mid \mbz_t)$ with a Gaussian is not sufficient to obtain a closed-form for the noisy likelihood since $p(\mby \mid \mbz_0) = \mc{N}(\mc{A}(\mc{D}(\mbz_0)), \sigma_y^2 \mbf{I}),$ which depends on the nonlinear decoder $\mc{D}$. Thus, prior work has focused on two classes of solutions. Decode-and-guide methods such as LDPS \citep{rout2023solving} approximate $p(\mbz_0 \mid \mbz_t)$ as a Dirac delta around the posterior mean $\mbz_{0\mid t} \triangleq \E[\mbz_0 \mid \mbz_t]$, so the noisy likelihood becomes 
\begin{equation} \label{eq:ldps}
\log p(\mby \mid \mc{D}(\mbz_{0 \mid t})) \propto -\norm{\mby - \mc{A}(\mc{D}(\mbz_{0 \mid t}))}{2}^2.
\end{equation}
%
%LDPS \citep{rout2023solving} approximates 
%$\log p(\mby \mid \mbz_t) \approx \log p(\mby \mid \mc{D}(\mbz_{0 \mid t})) \propto \norm{\mby - \mc{A}(\mc{D}(\mbz_{0 \mid t}))}{2}^2$.
Extensions iteratively optimize this residual \citep{song2023solving} or add regularization to the above approximation, such as ensuring $\mbz_{0 \mid t}$ is a 
fixed point of the autoencoder \citep{rout2023solving}. Attempts to model the uncertainty in LDM guidance have focused on adding computationally expensive second-order correction terms to this guidance update \citep{rout2024beyond}. As is shown in Figure~\ref{fig:visual_descrip}, decode-and-guide methods (red arrows) rely on decoder evaluations, eroding LDM efficiency gains. 

% To combat the problem of propagating uncertainty through the nonlinear decoder 
% The pathways from latent to pixel space are shown visually in Figure~\ref{fig:visual_descrip},
% where we see that 
% Furthermore,  Figure~\ref{fig:visual_descrip} shows that
% estimating $\Cov(\mby \mid \mbz_t)$ as a measure of uncertainty
% involves propagating uncertainty from $\mbz_t \to \mbz_0 \to \mbx_0 \to \mby$.
% While numerous methods can propagate
% uncertainty from $\mbz_t \to \mbz_0$, leveraging
% higher-order Tweedie's formulas \citep{boys2023tweedie,rissanen2024free}, uncertainty propagation through the nonlinear decoder
% from $\mbz_0 \to \mbx_0$ does not admit a closed-form solution.
% As a result, existing uncertainty-aware LDM guidance
% algorithms add computationally expensive second-order correction terms to guidance updates
% \citep{rout2024beyond}.

% {\color{red}In a latent diffusion model, why is the computation of $p(y \mid z_t)$ intractable? First, the computation of $p(y\mid z_0)$ seems tractable. Specifically, I think $p(y \mid x_0) \sim \mathcal{N}(\mathcal{A}(x_0);\sigma^2 I)$ and $x_0 = \mathcal{D}(z_0)$, so $p(y \mid z_0) \sim \mathcal{N}(\mathcal{A}(\mathcal{D}(z_0));\sigma^2 I)$. Then, $p(y \mid z_t) = \int_{z_0} p(y \mid z_0) p(z_0 \mid z_t) dz_0$. It should probably be said that the issue is this integral, and that LDPS avoids by using the mode of the distribution.}

To overcome this issue, native-latent methods such as SILO \citep{raphaeli2025silo} seek to replace the measurement operator $\mc{A}$, which maps the clean image $\mbx_0$ to the measurements $\mby$, by a native-latent operator $H$ that maps the latent image $\mc{E}(\mbx_0)$ to the latent measurements $\mc{E}(\mby)$. This is done by training a deep network $H_\theta$ to minimize a loss between $H_\theta(\mbz_{0\mid t})$ and latent measurements $\mc{E}(\mby)$. At inference time, the noisy likelihood score is approximated as in Equation~\ref{eq:ldps}, with $\mc{E}(\mby)$ replacing $\mby$ and the learned $H_\theta$ replacing $\mc{A} \circ \mc{D}$. However, since the true noisy likelihood involves an 
expectation over the conditional distribution $p(\mbz_0 \mid \mbz_t)$, which is highly anisotropic and input-dependent, 
collapsing the noisy likelihood score into the gradient of a loss from a single deterministic mapping ignores its uncertainty structure. In contrast to pixel-space guidance algorithms, decode-and-guide methods, and native-latent algorithms, \ours{} proposes an explicit probabilistic framework for approximating the noisy likelihood score and models the corresponding uncertainty.

\section{Anisotropic Native-Latent Operator Guidance}
\label{sec:method}

In this section, we present \ours{}, our framework for efficient uncertainty-aware guidance for LDMs. Our framework is built on two precise approximations, described below, which we later prove generalize implicit assumptions made in prior work~\citep{raphaeli2025silo}.
%Our framework rests on two precise approximations, which we describe below and afterwards show are generalizations of implicit assumptions made in prior work \citep{raphaeli2025silo}. 

\begin{approximation}[Sufficiency of Encoded Measurements]\label{apprx:suff}
    We replace conditioning on the raw measurement $\mby \in \R^n$ with
    conditioning on its encoding $\mc{E}(\mby) \in \R^k$:
    \begin{equation}\label{eq:approx1}
        p(\mbz_t \mid \mby) \;\approx\; p(\mbz_t \mid \mc{E}(\mby)).
    \end{equation}
\end{approximation}

This replacement is exact when $\mc{E}(\mby)$ is a sufficient statistic for $\mby$ with respect to the latent posterior.
In Section~\ref{sec:theory}, we prove in certain settings that common image processing tasks
like deblurring and super-resolution satisfy Approximation~\ref{apprx:suff}. Our next approximation is of the latent forward model.

\begin{approximation}[Gaussian Latent Forward Model]\label{apprx:gauss}
    We approximate the intractable latent forward model with a learned Gaussian:
    \begin{equation}\label{eq:obs_model}
        p(\mc{E}(\mby) \mid \mbz_t)
        \;\approx\; p_\theta(\mc{E}(\mby) \mid \mbz_t) \triangleq 
        \mc{N}\!\bigl(\mc{E}(\mby);\;
        \mu_\theta(\mbz_t, t, \sigma_y),\;
        \Sigma_\theta(\mbz_t, t, \sigma_y)\bigr),
    \end{equation}
    where $\mu_\theta$ and $\Sigma_\theta$ are learned mean and covariance functions\footnote{Henceforth, we will drop the function arguments $\mbz_t, t, \sigma_y$ for notational convenience}.
\end{approximation}

While the SILO guidance rule was derived without an explicit probabilistic formulation
(see Appendix \ref{app:silo} for details),
we take an alternate route and directly model the latent 
likelihood $p(\mc{E}(\mby) \mid \mbz_t)$. We further show in Appendix \ref{app:silo} how these two approximations can be viewed as relaxations of two implicit assumptions made in SILO \citep{raphaeli2025silo},
enabling a broader framework. 
Crucially, SILO corresponds to the special time and input-independent case $\Sigma_\theta \equiv \sigma^2 \mbf{I}$.
This is a strong structural assumption. The latent likelihood
$p(\mc{E}(\mby) \mid \mbz_t)$ depends on a nonlinear encoder and a complex denoising
process; different latent coordinates encode different spatial
structures, so their uncertainties given $\mbz_t$ should vary across both coordinates
and timesteps. Indeed, we show in Section~\ref{sec:theory} that even in an idealized 
linear Gaussian data model, the true covariance of $\mc{E}(\mby) \mid \mbz_t$ is inherently anisotropic
(Proposition~\ref{prop:sufficiency}, Part~\ref{prop:latent_lik}). Our framework generalizes SILO to allow for input-dependent, time-dependent,
and anisotropic covariance.

\paragraph{Training and Inference Algorithms.} Due to our explicit approximations, we can develop a simple, principled training objective 
and an uncertainty-aware guidance algorithm. Given the Gaussian model~\eqref{eq:obs_model}, we can use
maximum likelihood estimation to learn the mean and covariance by minimizing the
negative log-likelihood (NLL) \citep{bishop2006pattern, seitzer2022pitfalls}.
The per-sample NLL is
\begin{equation}\label{eq:nll}
    \ell_\theta(\mc{E}(\mby), \mbz_t, t, \sigma_y)
    \;=\;
    \frac{1}{2}\log|\Sigma_\theta|
    \;+\;
    \frac{1}{2}(\mc{E}(\mby) - \mu_\theta)^\top \Sigma_\theta^{-1} (\mc{E}(\mby) - \mu_\theta),
\end{equation}
and the training objective is
$\mc{L}_{\mathrm{NLL}}(\theta)
= \E_{\mbx, \bm{\epsilon}, t, \mbz_t}
[\ell_\theta(\mc{E}(\mby), \mbz_t, t, \sigma_y)]$.
At the optimum, the predicted covariance balances residual magnitude against uncertainty calibration, 
allowing the model to learn heterogeneous uncertainty estimates across coordinates, timesteps, and noise levels. 
In practice, we optimize $\mc{L}_{\mathrm{NLL}}$ in two stages: first learning a good mean prediction
with a frozen isotropic covariance, and then learning covariance structure on top of a fixed mean prediction (details in Appendix~\ref{app:exp_details}).

At sampling time, the guidance signal is
the score of the learned likelihood~\eqref{eq:obs_model}:
\begin{equation}\label{eq:guidance}
    \nabla_{\mbz_t} \log p_\theta(\mc{E}(\mby) \mid \mbz_t, t, \sigma_y)
    \;=\;
    -\frac{1}{2}\nabla_{\mbz_t}\log|\Sigma_\theta|
    \;-\;
    \frac{1}{2}\nabla_{\mbz_t}\bigl[
    (\mc{E}(\mby) - \mu_\theta)^\top \Sigma_\theta^{-1} (\mc{E}(\mby) - \mu_\theta)
    \bigr].
\end{equation}
This is a \emph{Mahalanobis-weighted} consistency signal:
$\Sigma_\theta^{-1}$ acts as a precision matrix that steers each latent coordinate
proportionally to the network's confidence in its prediction.

\paragraph{Practical Implementation.} While our framework allows for expressive covariance
parameterizations, storing and inverting large covariance matrices are often infeasible
in practice \citep{rissanen2024free}. In addition, differentiating through the covariance head adds memory and computational cost and can destabilize guidance, similar to observations in prior work \citep{rissanen2024free}. Thus, we parameterize the covariance
as a diagonal matrix (in a basis such as the Discrete Cosine Transform basis~\citep{ahmed1974discrete}) and use a stop-gradient approximation for $\Sigma_\theta$, so gradients only flow through the mean approximation. This yields efficient
training and inference, summarized in Algorithms \ref{alg:training} and \ref{alg:inference} in Appendix~\ref{app:algs}. We examine relaxing these approximations in Appendix~\ref{app:block_diag} and \ref{app:further_ablations}.
\section{Theoretical Results}
\label{sec:theory}

In this section, we consider a setting where  the data distribution follows a power law in the frequency domain, mimicking the spectral properties of natural image data \citep{van1996modelling}. We then demonstrate that in a linear model setting, deconvolution inverse problems satisfy Approximations~\ref{apprx:suff} and~\ref{apprx:gauss} (Proposition~\ref{prop:sufficiency}). Further, we characterize the \emph{anisotropic} and \emph{time-dependent} structure of the latent likelihood covariance. Our main result is to show a provable separation between existing isotropic likelihood approximations \citep{raphaeli2025silo}, and \ours{}, which converges to the true likelihood at a standard parametric rate (Theorem \ref{thm:scalar_approx}). In Section~\ref{sec:experiments}, we show that these theoretical ideas extend to general settings, and modeling covariance structure is helpful for large-scale nonlinear LDMs beyond deconvolution problems. All proofs are given in Appendix~\ref{app:proofs}.

\subsection{Problem Setup}

\paragraph{Notation.}
Let $\mbf{F}_d$ denote the $d$-dimensional Discrete
Fourier Transform matrix. 
We will drop the subscript $d$
when the dimension is clear from context. Assuming
the DC frequency is at the center of the DFT spectrum,
we index the Fourier-domain
by centered frequencies over the index set
\begin{equation}
    \Omega_d =
    \left\{
    -\tfrac{d-1}{2},\ldots,-1,0,1,\ldots,\tfrac{d-1}{2}
    \right\},
\end{equation}
where $d$ is assumed to be odd for notational convenience. 
We use $\Mconv$ to denote circular convolution with
stride $M$ and $\MconvT$ as transposed convolution with
stride $M$. We let $\mc{E}_\sharp p$ denote the
pushforward measure of $p$ under the map $\mc{E}$.
We consider the variance-preserving noising process for
the latent diffusion model with $T$ steps such that 
\begin{equation}
\mbz_t = \sqrt{\bar{\alpha}_t} \mbz_0
+ \sqrt{1 - \bar{\alpha}_t} \bm{\eps},
\quad \bm{\eps} \sim \mc{N}(\mbf{0}, \mbf{I}_k).
\end{equation}
Throughout this section, $\mbz_{0\mid t}$ denotes the
exact posterior mean $\E[\mbz_0 \mid \mbz_t]$ under the
idealized Gaussian model analyzed here. In practice,
LDMs are trained to predict this conditional mean
(e.g., via Tweedie's formula \citep{efron2011tweedie}),
and we do not study here the training and optimization
error of learning. Next, we state our assumptions.

\begin{assumption} \label{ass:data}
(Fourier-Diagonal Data Distribution)
There exist constants $\alpha>1$, $C > 0$ such that the clean image satisfies
$\mbx_0 \sim \mc{N}(\mbf{0}, \bm{\Sigma}_f)$ where
\begin{equation}
\bm{\Sigma}_f = \mbf{F}_d^*\,
\operatorname{diag}_{\omega\in\Omega_d}(\lambda_\omega)\,
\mbf{F}_d, \qquad \lambda_\omega = C(1+|\omega|)^{-\alpha}.
\end{equation}
\end{assumption}

%Note that since $\lambda_\omega = \lambda_{-\omega}$, the
%covariance $\bm{\Sigma}_f$ is real-valued, symmetric,
%$positive definite, and circulant.
Assumption~\ref{ass:data} is a standard assumption used in prior
work \citep{thaker2025frequency} and mimics the spectral
properties of natural image data, which are observed to
be approximately stationary in the frequency domain with a power law
spectrum \citep{van1996modelling}. Next, inspired by the strided convolutional structure of pretrained LDM autoencoders \citep{rombach2022high}, we assume a
similar convolutional structure for a linear LDM autoencoder.

\begin{assumption} \label{ass:conv_autoenc}
($M$-Strided Convolutional Autoencoder)
Assume $d \bmod M = 0$ and let
$k \triangleq d/M$ be the latent dimension of the
autoencoder. Suppose the encoder
$\mc{E}: \R^d \to \R^{k}$ and decoder
$\mc{D}: \R^{k} \to \R^d$ are linear maps of the form:
\begin{align}
    \mc{E}(\mbx) &= \mbf{W} (\mbf{g} \Mconv \mbx) \\
    \mc{D}(\mbz) &= \mbf{h} \MconvT (\mbf{Vz})
\end{align}
with $\mbf{g}, \mbf{h} \in \R^d$ and
$\mbf{W}, \mbf{V} \in \R^{k \times k}$.
\end{assumption}

We study learning this autoencoder with a 
regularized autoencoder objective \citep{tolstikhin2017wasserstein}, 
similar to the objective studied in~\citet{rout2023solving}:
\begin{equation} \label{eq:vae_loss}
    \min_{\mc{E}, \mc{D}} \mc{L}
    = \E\!\left[\norm{\mc{D}(\mc{E}(\mbx_0))
    - \mbx_0}{2}^2\right]
    + \gamma\, \KL{\mc{E}_\sharp p}
    {\mc{N}(0,\mbf{I}_k)}.
\end{equation}
Finally, we study deconvolution problems, which covers common image restoration tasks such as deblurring.
\begin{assumption} \label{ass:ip}
(Deconvolution Inverse Problem)
Assume a measurement model $\mby = \mbA \mbx_0 + \sigma_y \bm{\epsilon}$, where $\sigma_y > 0$, $\bm{\epsilon} \sim \mc{N}(\bm{0}, \mbf{I})$, and $\mbA\mbx = \bm{\psi} \circledast \mbx$ is a circular convolution. Let $a_\omega = \text{diag}(\mbf{F} \mbA \mbf{F}^*)_\omega$ be the frequency response of $\mbA$ at frequency $\omega$. 
\end{assumption}

%\begin{remark}
%    In practice, convolutional autoencoders use a kernel
%    of size $L \ll d$ that slides over the input, rather
%    than a circular kernel of size $d$. This restricts
%    the filter to $L$ degrees of freedom in the frequency
%    domain, so it can only approximate the ideal low-pass
%    response, introducing additional aliasing that
%    vanishes as $L \to d$. Since this is a standard
%    Finite Impulse Response (FIR) approximation artifact
%    orthogonal to our analysis, we work with $L = d$
%    throughout.
%\end{remark}

%
% This is not the standard VAE
% ELBO \citep{kingma2013auto}: it regularizes the
% aggregated encoded distribution $\mc{E}_\sharp p$,
% rather than a per-example approximate posterior
% $q(\mbz\mid \mbx)$. We use this prior-regularized
% autoencoder objective \citep{tolstikhin2017wasserstein}
% only as an analytically tractable model of an encoder,
% not as a claim about the exact training objective of
% the pretrained LDM autoencoder.

\subsection{Verifying Approximation \ref{apprx:suff}
and \ref{apprx:gauss}}

We first show that Approximations \ref{apprx:suff} and \ref{apprx:gauss} are exact under our model assumptions, i.e., there is no approximation error. Before  studying the approximations, we characterize the structure of the latent space by characterizing the optimal encoder and decoder under the objective given in Equation~\eqref{eq:vae_loss}.
\begin{proposition} \label{prop:autoenc_optim}
%    The following parameter setting of the $M$-strided convolutional autoencoder is a global minimizer of the loss given in Equation \eqref{eq:vae_loss}.
%    In centered Fourier coordinates, the analysis and synthesis filters are ideal low-pass filters:
The parameters of an $M$-strided convolutional autoencoder that achieve the global minimum of the loss in~\eqref{eq:vae_loss} are given by:
\begin{equation}
(\mbf{F}_d \mbf{g})_\omega = (\mbf{F}_d \mbf{h})_\omega =
\begin{cases}
\sqrt{M} & \text{for } \omega \in \Omega_k,\\
0 & \text{for } \omega \notin \Omega_k,
\end{cases}
\qquad
\begin{aligned}
\mbf{W} &=
\mbf{F}_k\,
\operatorname{diag}_{\omega\in\Omega_k}
(\lambda_\omega^{-1/2})\,
\mbf{F}_k^*,
\\
\mbf{V} &=
\mbf{F}_k\,
\operatorname{diag}_{\omega\in\Omega_k}
(\lambda_\omega^{1/2})\,
\mbf{F}_k^* .
\end{aligned}
\end{equation}
    % %
    % The latent-space linear maps whiten and unwhiten
    % the retained frequencies:
    % %
    % \begin{align}
    % \mbf{W} &= \mbf{F}_k\,
    % \operatorname{diag}_{\omega\in\Omega_k}
    % (\lambda_\omega^{-1/2})\,
    % \mbf{F}_k^*, \label{eq:W_opt} \\[4pt]
    % \mbf{V} &= \mbf{F}_k\,
    % \operatorname{diag}_{\omega\in\Omega_k}
    % (\lambda_\omega^{1/2})\,
    % \mbf{F}_k^*. \label{eq:V_opt}
    % \end{align}
\end{proposition}

%The proof is given in Appendix~\ref{app:proof_autoenc}.
This proposition states the optimal parameters of the
convolutional encoder correspond to an ideal low-pass
filter that retains the $k$ lowest-magnitude centered
frequencies $\Omega_k$ and discards all frequencies in
$\Omega_d \setminus \Omega_k$.
%While our characterization assumes linearity, this
%low-pass structure has also been enforced via additional
%regularization in certain nonlinear LDM
%autoencoders \citep{skorokhodov2025improving}.
For the remainder of this section, we instantiate
$(\mc E,\mc D)$ with the optimal parameters
characterized in Proposition~\ref{prop:autoenc_optim}. We now present
the main result of this subsection.

%We now justify Approximations \ref{apprx:suff}
%and \ref{apprx:gauss} under our model.
%
\begin{proposition}[Approximations \ref{apprx:suff}
and \ref{apprx:gauss}
are exact]\label{prop:sufficiency}
    Under Assumptions~\ref{ass:data}--\ref{ass:ip} with
    the optimal autoencoder
    (Proposition~\ref{prop:autoenc_optim}), the following
    hold:
    \begin{enumerate}[leftmargin=*]
        \item \textbf{(Sufficiency.)}
        $\mc{E}(\mby)$ is sufficient for predicting $\mbz_t$:
        \begin{equation}
        p(\mbz_t \mid \mby)
        = p(\mbz_t \mid \mc{E}(\mby)).
        \end{equation}

        \item \label{prop:latent_lik}
        \textbf{(Gaussian likelihood.)}
        The true $p(\mc{E}(\mby) \mid \mbz_t)$ is a
        Gaussian distribution with mean $\bm{\mu}_t$ and
        covariance $\mbf{C}_t$ given by:
        \begin{equation}
            \bm{\mu}_t = \mbf{F}_k^*\,
            \operatorname{diag}_{\omega\in\Omega_k}
            (a_\omega)\, \mbf{F}_k \mbz_{0 \mid t}, \qquad  \mbf{C}_t = \mbf{F}_k^*\,
            \operatorname{diag}_{\omega\in\Omega_k}
            \bigl(c_\omega(t)\bigr)\, \mbf{F}_k,
        \end{equation}
        where $\{a_\omega\}_{\omega \in \Omega_k}$ denotes
        the frequency response of $\mbA$ for the retained
        frequencies and
        \begin{equation}\label{eq:cj}
            c_\omega(t) =
            (1 - \bar{\alpha}_t)|a_\omega|^2
            + \frac{\sigma_y^2}{\lambda_\omega},
            \qquad \omega \in \Omega_k.
        \end{equation}
    \end{enumerate}
\end{proposition}
%
%The proof is given in Appendix~\ref{app:proof_sufficiency}. 
Part~1 proves that encoding the measurement preserves exactly the information needed for the latent posterior, motivating native-latent guidance. Part~2 shows the Gaussian structure of the latent likelihood induced by the data frequency structure. The proof of Part~2 draws a connection between native-latent guidance and multirate signal processing, specifically utilizing the Noble identities \citep{vaidyanathan2006multirate} to obtain a closed form for the likelihood. The covariance eigenvalues $c_\omega(t)$ consist of two terms: $(1 - \bar\alpha_t)|a_\omega|^2$, which captures diffusion uncertainty propagated through the forward operator, and $\sigma_y^2/\lambda_\omega$, which captures measurement noise amplified by the encoder's whitening. Crucially, both terms are frequency-dependent, making the covariance inherently anisotropic. 

\subsection{Demonstrating Provable Need for Covariance
Modeling}

Our main result for our setting is that anisotropic, uncertainty-aware guidance is necessary for correct sampling, whereas isotropic guidance incurs a strictly positive error relative to the true likelihood.

%%% --- Theorem: Scalar Covariance Approx Error --- %%%

\begin{theorem}[Provable Separation of Isotropic vs.\
Anisotropic Covariance]\label{thm:scalar_approx}
    Under Assumptions~\ref{ass:data}--\ref{ass:ip} with
    the optimal autoencoder
    (Proposition~\ref{prop:autoenc_optim}),
    let $\mbf{C}_t$ be the true likelihood covariance
    from~\eqref{eq:cj} with eigenvalues
    $c_\omega(t)$.
    Then, the following statements hold:
    \begin{enumerate}[leftmargin=*]
        \item \textbf{(SILO: irreducible error.)}
        For any scalar $\eta > 0$, define the isotropic
        approximation
        $\tilde{p}(\mc{E}(\mby) \mid \mbz_t)
        \coloneqq \mc{N}(\bm{\mu}_t,\,
        \eta^2\,\mbf{I}_k)$.
        Then
        \begin{equation}\label{eq:kl_amgm}
            \KL{p(\mc{E}(\mby) \mid \mbz_t)}
                {\tilde{p}(\mc{E}(\mby) \mid \mbz_t)}
            \geq \frac{k}{2}\,\log
            \frac{\mathrm{AM}\bigl(
            \{c_\omega(t)\}_{\omega\in\Omega_k}\bigr)}
                 {\mathrm{GM}\bigl(
            \{c_\omega(t)\}_{\omega\in\Omega_k}\bigr)},
        \end{equation}
        where $\mathrm{AM}$ and $\mathrm{GM}$ are the
        arithmetic and geometric means respectively. This
        lower bound is attained at
        $\eta^{2} = \mathrm{AM}\!\bigl(
        \{c_\omega(t)\}_{\omega\in\Omega_k}\bigr)$
        and under Assumptions~\ref{ass:data}--\ref{ass:ip}, it is strictly positive 
        with probability $1$ for $t \sim \mc{U}[0, T]$.

        \item \textbf{(\ours: convergence to true
        likelihood.)}
        Let $\hat{p}$ denote the MLE of the Gaussian
        model~\eqref{eq:obs_model} with linear mean parameterization and Fourier-diagonal
        covariance, trained on
        $n$ i.i.d.\ samples at a fixed timestep $t$.
        Then
        \begin{equation}\label{eq:mle_rate}
            \E\!\left[
            \KL{p(\mc{E}(\mby) \mid \mbz_t)}
            {\hat{p}(\mc{E}(\mby) \mid \mbz_t)}
            \right]
            \;=\; O\!\left(\frac{k}{n}\right).
        \end{equation}
    \end{enumerate}
\end{theorem}

\begin{wrapfigure}{r}{0.5\linewidth}
    \centering
    \vspace{-4mm}
    \subfloat[Covariance eigenvalues $c_\omega(t)$ at
    different diffusion timesteps
    ($\alpha{=}2.5$).\label{fig:cov_spectrum}]{%
        \includegraphics[width=0.48\linewidth]
        {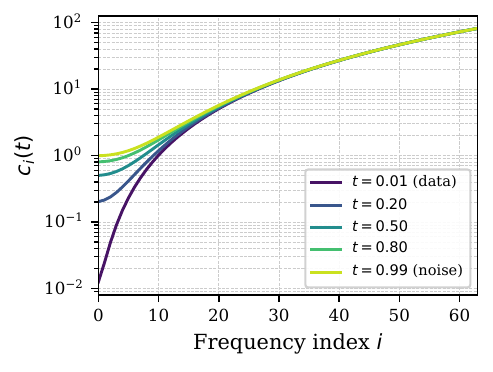}} 
    \hspace{0.5em}
    \subfloat[KL lower bound~\eqref{eq:kl_amgm} vs.\
    diffusion time $t$ for different spectral decays
    $\alpha$.\label{fig:kl_bound}]{%
        \includegraphics[width=0.48\linewidth]
        {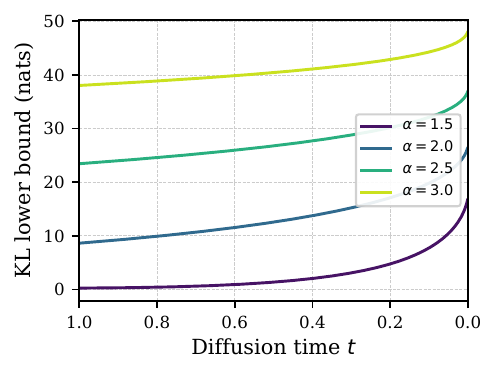}}
    \caption{Visualizing our theorem bounds.}
    \label{fig:theory_plots}
    \vspace{-4mm}
\end{wrapfigure}

%The proof is given in
%Appendix~\ref{app:proof_scalar_approx}.
\paragraph{Implications.}
Theorem~\ref{thm:scalar_approx} gives a clean
separation within this model: an isotropic
covariance family has an irreducible KL gap, while an
anisotropic family 
recovers the true likelihood at the standard parametric
rate.
Figure~\ref{fig:theory_plots} illustrates the 
qualitative behavior of the covariance eigenvalues and
the KL lower bound. First,
Figure~\ref{fig:theory_plots}a shows that the covariance
spectrum is highly anisotropic, especially near the end
of diffusion sampling. Second,
Figure~\ref{fig:theory_plots}b shows the KL penalty
grows with the spectral decay rate $\alpha$. Crucially,
the isotropic approximation error is largest at the
end of the reverse process, where
guidance precision matters most for preserving 
high-frequency
details \citep{rissanen2022generative}. This may provide
theoretical evidence as to why SILO produces
smoothed-out reconstructions
(Figure \ref{fig:teaser_qual}).
More analysis of this lower bound is given in
Appendix~\ref{app:theory_analysis}.

\section{Experiments}
\label{sec:experiments}

\subsection{Experimental Setup}

\paragraph{Tasks, Datasets, and Models.} We consider five inverse
problems, grouped into three deconvolution tasks (Gaussian blur,
4x and 8x super-resolution) and two general inverse
problems (box inpainting, JPEG decompression). The measurement
noise for each task is $\sigma_y = 0.02$. Further task details
are provided in Appendix \ref{app:exp_details}. We evaluate on
a fixed 1000 image test set from the FFHQ dataset \citep{Karras2019ASG} and the
ImageNet-1000 dataset \citep{deng2009imagenet}, both at
$512\times512$ resolution. We utilize the
pretrained LDM  
Realistic Vision-v5.1 \citep{realisticvision_v51} from HuggingFace.
For all methods, we utilize 100 sampling steps for the FFHQ dataset and 250 sampling
steps for the more complex ImageNet dataset.

\paragraph{Baselines and Metrics.} We compare to four
leading LDM inverse problem solvers. The first 
class is decode-and-guide methods: LDPS \citep{rout2023solving}, 
PSLD \citep{rout2023solving}, and
STSL \citep{rout2024beyond}. The other class
is native-latent guidance, such as SILO~\citep{raphaeli2025silo}. 
These baselines serve different roles. SILO is the closest
comparison because it is also a trained native-latent method.
LDPS and PSLD are representative
decode-and-guide posterior samplers that do not use the same
native-latent operator, and STSL represents a more
expensive uncertainty-weighted decode-and-guide approach.
We report FID and LPIPS for perceptual quality~\citep{zhang2018unreasonable},
and PSNR for distortion, each with subscripts denoting the standard error of the mean
over the 1000 test examples. We also report sampling time in seconds
per image.

\paragraph{Details of \ours.} We train a lightweight variance head alongside the mean prediction network, adding only $\approx 3.4\%$ additional parameters (480K out of 14.2M). The variance head takes inspiration from the Readout-Guidance architecture \citep{luo2024readout} and takes in as input aggregated features from the LDM UNet denoiser and a $\sigma_y$ embedding. It then outputs a log-variance map in latent space, parametrized as diagonal both in spatial and DCT domain, with the latter inspired by our theoretical results. We test richer parameterizations of the covariance in Appendix \ref{app:block_diag}. We denote \ours{} and \ours-DCT as our method with the spatial and DCT covariances respectively. The exact architecture and training details are in
Appendix~\ref{app:exp_details}.

\subsection{Main Results}

Tables~\ref{tab:ffhq_deconv} and~\ref{tab:ffhq_general} show that, for both datasets, \ours{} consistently improves perceptual metrics over the native-latent baseline, SILO, on most tasks while maintaining a nearly identical inference time. This highlights our main conceptual message: modeling covariance structure is critical for native-latent guidance. Modeling the covariance in the DCT domain rather than in spatial domain yields further improvements in most tasks, especially for JPEG decompression, where the forward operator also operates in the DCT domain \citep{wallace1991jpeg}. Compared with decode-and-guide baselines, \ours{} generally provides a better quality-efficiency tradeoff, matching or exceeding these baselines on the FFHQ dataset. Notably, \ours{} is $18\times$ faster than STSL. On ImageNet, \ours{} also yields substantial gains over SILO (e.g., $62.5 \to 26.9$ FID on Gaussian deblurring, $132.1 \to 71.1$ on 4× super-resolution), confirming that covariance modeling transfers beyond face datasets. A gap to decode-and-guide baselines remains on super-resolution; since both native-latent methods show degraded PSNR on ImageNet relative to FFHQ, we attribute this to difficulty of mean prediction rather than covariance. Improving $\mu_\theta$ (e.g., by incorporating class labels as conditional inputs) is orthogonal to our contribution.
%
%On the more diverse ImageNet dataset, a performance gap remains between certain decode-and-guide baselines and native-latent approaches for super-resolution tasks. We attribute this gap to difficulty in training the mean network $\mu_\theta$ as both SILO and \ours{} achieve lower PSNR values on ImageNet than on FFHQ. As this work focuses primarily on covariance structure, where \ours{} significantly improves upon SILO even on ImageNet, we leave further optimization of the mean network for future work (e.g., incorporating class labels as conditional inputs).
%
Finally, results on inpainting and JPEG decompression (Table~\ref{tab:ffhq_general}) suggest that, even when the exact sufficiency argument from Section~\ref{sec:theory} no longer applies, native-latent guidance remains empirically effective and continues to benefit from covariance-aware weighting.

\begin{figure*}[ht]
    \centering
    \includegraphics[width=0.85\linewidth]{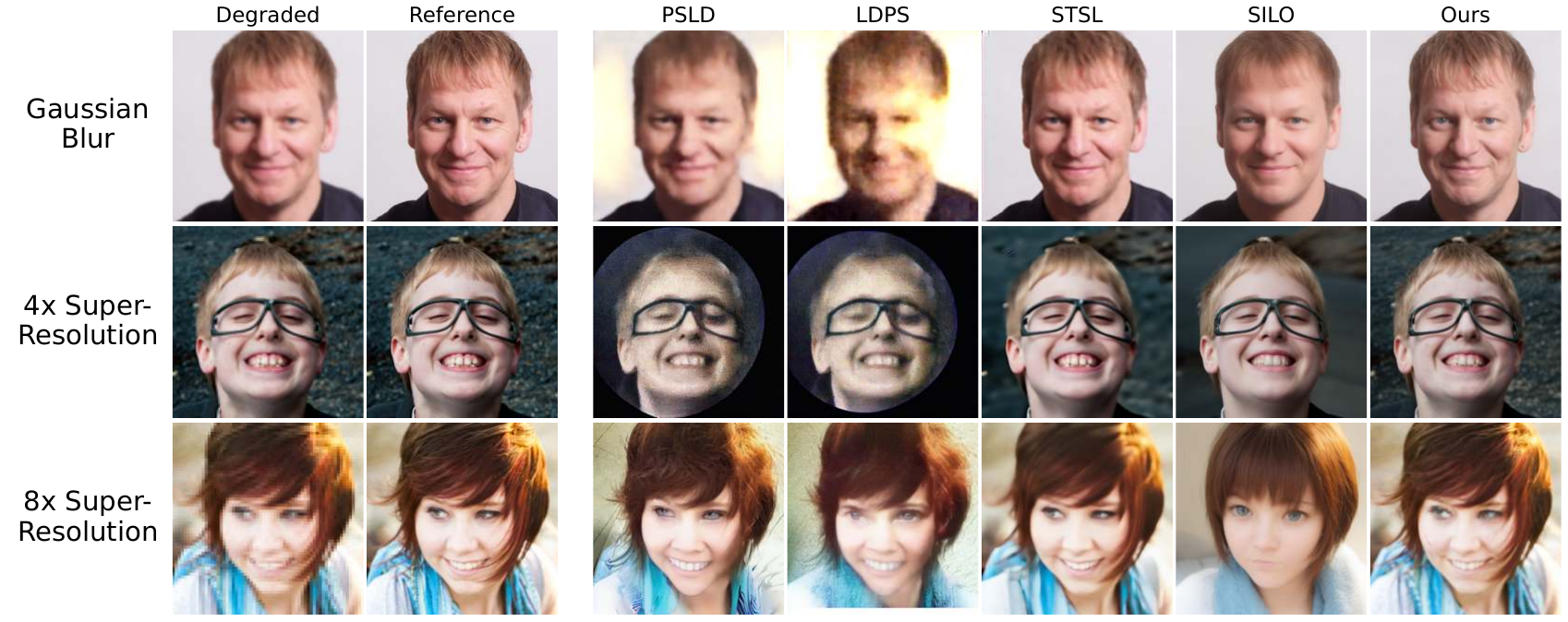}
    \caption{Qualitative comparison on FFHQ deconvolution
    tasks. Best viewed zoomed in.}
    \label{fig:ffhq_qual}
\end{figure*}

Figure~\ref{fig:ffhq_qual} illustrates the qualitative differences among methods. The isotropic covariance in SILO leads to over-smoothed reconstructions that lack high-frequency detail. STSL recovers finer detail but can also introduce some visual artifacts. In contrast, \ours{} produces visually plausible and consistent reconstructions. Additional qualitative results are provided in Appendix~\ref{app:qual}.

Figure~\ref{fig:variance_maps} visualizes the learned log-variance spatial maps for a single input across four tasks at two different timesteps. It highlights that the learned covariances are anisotropic, input- and time-dependent. Across different tasks, the variance captures operator-specific structure; e.g., in super-resolution it concentrates along edges and fine textures, where high-frequency information is lost. At early timesteps, the variance is relatively uniform, whereas at later timesteps, spatial structure emerges and becomes more pronounced. The learned covariance is also well-calibrated on held-out data across all tasks as shown in Appendix~\ref{app:calibration}.

\newcommand{\sem}[1]{{\scriptsize$_{(#1)}$}}
\newcommand{\g}{\cellcolor{gray!10}}

\begin{figure}[t!]
    \centering
    \subfloat[Learned spatial variance maps. Variance is averaged across the
    4 latent channels.\label{fig:variance_maps}]{%
        \raisebox{-.5\height}{\includegraphics[width=0.37\linewidth]{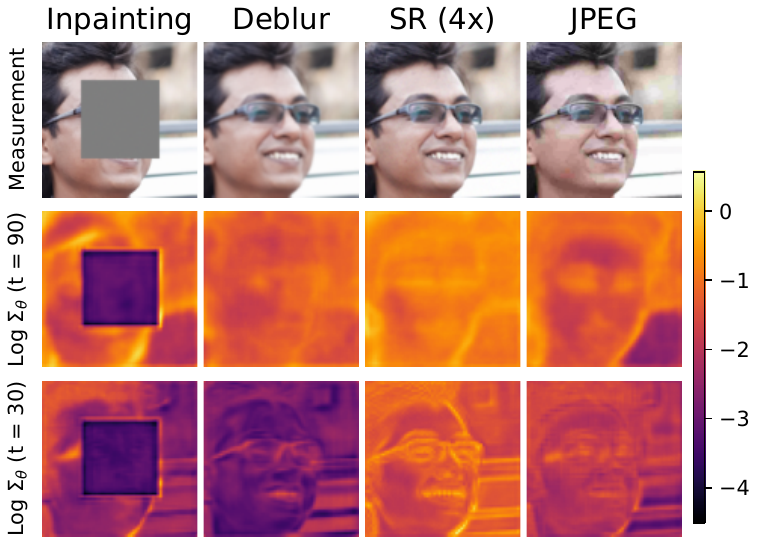}}}
    \hfill
    \subfloat[Ablation of covariance structure on Gaussian deblurring (FFHQ).\label{tab:ablation_variance}]{%
        \raisebox{-.5\height}{\footnotesize
        \begin{tabular}{l|ccc}
        \toprule
        Variant
          & FID$\downarrow$ & LPIPS$\downarrow$
          & PSNR$\uparrow$ \\
        \midrule
        Isotropic (SILO)
          & 47.3 & 0.367\sem{.002}
          & 25.03\sem{.06} \\
        Time-independent
          & \underline{41.09} & 0.378\sem{.002} %0.302\std{.10}
          & 25.48\sem{.09} \\
        Theory-predicted
          & 48.87 & \underline{0.353}\sem{.001}
          & \underline{25.86}\sem{.05} \\
        \ours{} 
          & \textbf{29.79} & \textbf{0.326}\sem{.002}
          & \textbf{27.34}\sem{.09} \\
        %Ours (DCT)
       %   & \underline{30.9} & \textbf{0.190}\std{.06}
       %   & \underline{26.90}\std{2.0} \\
        \bottomrule
        \end{tabular}}}
    \caption{Learned variance maps and ablation of covariance structure. \textbf{(a)}~Task-dependent variance maps show the model learns distinct spatial patterns for each inverse problem. \textbf{(b)}~Progressively richer covariance models improve reconstruction quality.}
    \label{fig:ablation_combined}
\end{figure}

\begin{table*}[ht]
\centering
\footnotesize
\setlength{\tabcolsep}{2.5pt}
\resizebox{\textwidth}{!}{%
\begin{tabular}{cc l c ccc ccc ccc}
\toprule
& & & & \multicolumn{3}{c}{Gaussian Blur}
  & \multicolumn{3}{c}{4x Super-Resolution}
  & \multicolumn{3}{c}{8x Super-Resolution} \\
\cmidrule(lr){5-7} \cmidrule(lr){8-10} \cmidrule(lr){11-13}
& & Method & Time (s)
  & FID$\downarrow$ & LPIPS$\downarrow$ & PSNR$\uparrow$
  & FID$\downarrow$ & LPIPS$\downarrow$ & PSNR$\uparrow$
  & FID$\downarrow$ & LPIPS$\downarrow$ & PSNR$\uparrow$ \\
\midrule
\multirow{6}{*}{\rotatebox[origin=c]{90}{\textbf{FFHQ}}}
& \multirow{3}{*}{\rotatebox[origin=c]{90}{\scriptsize\textit{Decode}}}
& \g LDPS & \g \underline{52.2}
  & \g 84.8 & \g 0.564\sem{.004} & \g 21.48\sem{.16}
  & \g 68.6 & \g 0.554\sem{.003} & \g 20.70\sem{.11}
  & \g 55.5 & \g 0.536\sem{.003} & \g 21.24\sem{.08} \\
& & \g PSLD & \g 62.2
  & \g 70.8 & \g 0.535\sem{.003} & \g 22.65\sem{.14}
  & \g 53.5 & \g 0.505\sem{.003} & \g 22.52\sem{.10}
  & \g 54.0 & \g 0.513\sem{.003} & \g 21.68\sem{.08} \\
& & \g STSL & \g 433
  & \g 41.7 & \g 0.419\sem{.002} & \g \textbf{27.36}\sem{.07}
  & \g 39.7 & \g 0.415\sem{.002} & \g 26.88\sem{.10}
  & \g 38.6 & \g 0.439\sem{.002} & \g 25.58\sem{.08} \\
\cmidrule{2-13}
& \multirow{3}{*}{\rotatebox[origin=c]{90}{\scriptsize\textit{Native}}}
& SILO & \textbf{24.0}
  & 47.3 & 0.367\sem{.002} & 25.03\sem{.06}
  & \underline{38.8} & \underline{0.331}\sem{.002} & 26.26\sem{.06}
  & \underline{35.0} & \underline{0.356}\sem{.002} & \underline{25.81}\sem{.06} \\
% SILO: IP 86.64 & 0.439 & 19.16 & 0.608
& & \ours & \textbf{23.7}
  & \textbf{29.8} & \textbf{0.326}\sem{.002} & \textbf{27.34}\sem{.09}
  & \textbf{27.0} & \textbf{0.303}\sem{.002} & \textbf{27.56}\sem{.08}
  & 38.8 & 0.357\sem{.002} & 25.39\sem{.05} \\
& & \ours-DCT & \textbf{24.1}
  & \underline{30.9} & \underline{0.338}\sem{.002} & \underline{26.36}\sem{.10}
  & \textbf{26.4} & \textbf{0.302}\sem{.002} & \underline{26.92}\sem{.09}
  & \textbf{31.1} & \textbf{0.341}\sem{.002} & \textbf{26.37}\sem{.07} \\
\midrule[0.08em]
\multirow{6}{*}{\rotatebox[origin=c]{90}{\textbf{ImageNet}}}
& \multirow{3}{*}{\rotatebox[origin=c]{90}{\scriptsize\textit{Decode}}}
& \g LDPS & \g \underline{137.9}
  & \g 55.9 & \g 0.463\sem{.003} & \g 24.99\sem{.13}
  & \g 111.7 & \g 0.548\sem{.003} & \g 23.26\sem{.12}
  & \g 149.3 & \g 0.576\sem{.003} & \g \underline{21.58}\sem{.12} \\
& & \g PSLD & \g 164.8
  & \g 53.4 & \g 0.455\sem{.003} & \g \underline{25.05}\sem{.13}
  & \g 95.8 & \g 0.546\sem{.003} & \g \underline{23.28}\sem{.12}
  & \g 146.9 & \g 0.577\sem{.003} & \g 21.38\sem{.11} \\
& & \g STSL & \g 1155
  & \g 39.8 & \g 0.404\sem{.003} & \g \textbf{25.95}\sem{.13}
  & \g \textbf{38.2} & \g \textbf{0.422}\sem{.003} & \g \textbf{25.95}\sem{.13}
  & \g \textbf{69.6} & \g \textbf{0.491}\sem{.003} & \g \textbf{23.41}\sem{.12} \\
\cmidrule{2-13}
& \multirow{3}{*}{\rotatebox[origin=c]{90}{\scriptsize\textit{Native}}}
& SILO & \textbf{58.6}
  & 62.5 & 0.450\sem{.003} & 22.94\sem{.11}
  & 132.1 & 0.541\sem{.004} & 20.73\sem{.10}
  & 129.1 & 0.544\sem{.004} & 19.41\sem{.09} \\
& & \ours & \textbf{57.8}
  & \underline{28.6} & \textbf{0.370}\sem{.003} & 24.77\sem{.14}
  & 85.8 & 0.477\sem{.003} & 21.71\sem{.09}
  & 119.2 & 0.524\sem{.003} & 19.68\sem{.08} \\
& & \ours-DCT & \textbf{59.1}
  & \textbf{26.9} & \underline{0.403}\sem{.002} & 23.74\sem{.10}
  & \underline{71.1} & \underline{0.440}\sem{.002} & 22.44\sem{.08}
  & \underline{105.2} & \underline{0.509}\sem{.003} & 19.83\sem{.09} \\
\bottomrule
\end{tabular}}
\caption{Quantitative results on deconvolution tasks on
the FFHQ and ImageNet datasets. Bolded entries indicate best performance,
underlined entries indicate second best performance, and subscripts denote standard error of mean. We bold multiple rows 
in the event of approximately tied results. Gaussian deblurring is the representative
task for timings since timings do not vary significantly across the different forward operators.}
\label{tab:ffhq_deconv}
\end{table*}

\begin{table*}[ht]
\centering
\footnotesize
\setlength{\tabcolsep}{2.5pt}
\begin{tabular}{cc l c ccc ccc}
\toprule
& & & & \multicolumn{3}{c}{Box Inpainting}
  & \multicolumn{3}{c}{JPEG}  \\
\cmidrule(lr){5-7} \cmidrule(lr){8-10}
& & Method & Time (s)
  & FID$\downarrow$ & LPIPS$\downarrow$ & PSNR$\uparrow$
  & FID$\downarrow$ & LPIPS$\downarrow$ & PSNR$\uparrow$ \\
\midrule
\multirow{6}{*}{\rotatebox[origin=c]{90}{\textbf{FFHQ}}}
& \multirow{3}{*}{\rotatebox[origin=c]{90}{\scriptsize\textit{Decode}}}
& \g LDPS & \g \underline{52.2}
  & \g 68.2 & \g 0.205\sem{.001} & \g 16.61\sem{.12}
  & \g 101.0 & \g 0.612\sem{.003} & \g 17.83\sem{.14} \\
& & \g PSLD & \g 62.2
  & \g 62.7 & \g 0.196\sem{.001} & \g 17.09\sem{.12}
  & \g N/A & \g N/A & \g N/A \\
& & \g STSL & \g 433
  & \g 63.1 & \g 0.191\sem{.001} & \g 18.19\sem{.13}
  & \g 63.5 & \g 0.502\sem{.003} & \g 22.09\sem{.17} \\
\cmidrule{2-10}
& \multirow{3}{*}{\rotatebox[origin=c]{90}{\scriptsize\textit{Native}}}
& SILO & \textbf{24.0}
  & \underline{32.8} & \underline{0.142}\sem{.001} & \textbf{22.59}\sem{.08}
  & 68.3 & 0.471\sem{.002} & 20.87\sem{.06} \\
%Resample & & - & - & - & - & - & - \\
& & \ours & \textbf{23.7}
  & \textbf{22.8} & \textbf{0.135}\sem{.001} & 21.31\sem{.11}
  & \underline{46.3} & \underline{0.393}\sem{.002} & \underline{23.91}\sem{.06} \\
& & \ours-DCT & \textbf{24.1}
  & \textbf{22.1} & \textbf{0.132}\sem{.001} & \underline{22.08}\sem{.10}
  & \textbf{27.3} & \textbf{0.326}\sem{.002} & \textbf{26.26}\sem{.09} \\
\bottomrule
\end{tabular}
\caption{Quantitative results on center box inpainting and
JPEG decompression on the FFHQ dataset. Note PSLD does not
apply to nonlinear inverse problems.}
\label{tab:ffhq_general}
\end{table*}

\paragraph{Ablation Studies.}
We now ablate the covariance structure on a Gaussian deblurring task on the FFHQ dataset to support the need for anisotropic, input- and time-dependent covariance. We compare four variants:
(i)~\emph{isotropic covariance}, which corresponds to SILO and uses a scalar variance;
(ii)~\emph{time-independent covariance}, which computes the variance only at the first timestep and reuses it throughout sampling;
(iii)~\emph{theory-predicted covariance}, which uses the diagonal covariance predicted by Proposition~\ref{prop:sufficiency}, i.e., the optimal covariance under our assumptions; and
(iv)~our model with time- and input-dependent
covariance. Figure~\ref{tab:ablation_variance} shows that the theory-predicted covariance already provides improvements in perceptual quality, suggesting that the spectral covariance structure is beneficial. As we further incorporate input and time dependence, both perceptual and distortion metrics improve, highlighting the importance of modeling anisotropic, input- and time-dependent covariance. We prove additional ablation studies on varying guidance scale as well as considering richer covariance structure in Appendix~\ref{app:further_ablations}.

\section{Conclusion}

We introduced \ours{}, a framework for efficient uncertainty-aware guidance with pretrained latent diffusion models. By interpreting existing native-latent guidance as approximate inference under a Gaussian latent likelihood, we identify covariance modeling as the \emph{missing degree of freedom} in native-latent guidance methods. In a Fourier-diagonal data model, we proved that the true latent likelihood is inherently anisotropic, and any isotropic approximation incurs an irreducible KL gap, whereas \ours{} recovers the true likelihood at the standard parametric rate. Empirically, modeling the mean and covariance together yields consistent improvements in perceptual reconstruction quality, even with a lightweight diagonal covariance head. 

\paragraph{Limitations and Future Work.} \ours{} inherits some limitations of native-latent guidance, as discussed in detail in  Appendix~\ref{app:limitations}. Nonetheless, we hope the probabilistic formulation of native-latent guidance can inspire work on uncertainty-aware native-latent guidance, possibly considering richer covariance parameterizations. %See Appendix~\ref{app:limitations} for further discussion.

{
    \small
    \bibliographystyle{iclr2027_conference}
    \bibliography{main}
}

%%%%%%%%%%%%%%%%%%%%%%%%%%%%%%%%%%%%%%%%%%%%%%%%%%%%%%%%%%%%

\clearpage
\appendix
\numberwithin{equation}{section}
\setcounter{equation}{0}
% ============================================================
%  Appendix
% ============================================================

\renewcommand{\contentsname}{Appendix Table of Contents}
\startcontents[appendix]
\printcontents[appendix]{}{1}{\setcounter{tocdepth}{2}}
\vspace{1em}

\FloatBarrier
\section{Implicit Assumptions Behind SILO}
\label{app:silo}

The central object needed for posterior sampling with latent diffusion models is the
noisy likelihood score $\nabla_{\mbz_t} \log p(\mby \mid \mbz_t)$. The main challenge
is that $\mby$ is obtained by a forward model operating on images, whereas $\mbz_t$ 
is in the latent space of the LDM encoder.
Decode-and-guide methods~\citep{rout2023solving, rout2024beyond}
approximate this noisy likelihood score by mapping latents back to pixel space at each step,
which inevitably sacrifices the efficiency of operating in the latent domain.
SILO~\citep{raphaeli2025silo} takes a different approach: it encodes the measurement
via $\mc{E}(\mby)$ and directly learns a latent forward operator $H_\theta$. 
This latent operator is trained to map denoised latent estimates $\mbz_{0 \mid t}$ directly to
$\mc{E}(\mby)$ by minimizing
a residual between $H_\theta(\mbz_{0 \mid t})$ and $\mc{E}(\mby)$.
At inference time, the noisy likelihood score is approximated as 
the negative gradient of an $\ell_2$ loss:
\begin{equation}\label{eq:silo_loss}
    -\nabla_{\mbz_t} \norm{\mc{E}(\mby) - H_\theta(\mbz_{0\mid t})}{2}^2,
\end{equation}
While this yields fast and effective guidance, the distributional
assumptions that justify the training objective and justify ~\eqref{eq:silo_loss} as the correct 
guidance signal are left implicit.  For completeness, we repeat
the derivation of SILO native-latent guidance below. Assuming we
have a trained network $H_\theta$ that approximates the forward operator $\mc{A}$ in latent space (as shown in Figure~\ref{fig:visual_descrip}), the starting point
of the derivation is the pixel-space residual

\begin{align}
    \norm{\mby - \mc{A}(\mbx_0)}{2}^2 &\approx \norm{\mc{D}(\mc{E}(\mby)) - \mc{D}(\mc{E}(\mc{A}(\mbx_0)))}{2}^2 \\
    &\approx \norm{\mc{D}(\mc{E}(\mby)) - \mc{D}(H_\theta(\mbz_0))}{2}^2 \\
    &\leq C_{\mc{D}}^2 \norm{\mc{E}(\mby) - H_\theta(\mbz_0)}{2}^2
\end{align}

Above, $C_\mc{D}$ denotes the Lipschitz constant of the LDM decoder. The final equation is then used as a proxy to $\nabla_{\mbz_t} \log p(\mby \mid \mbz_t)$. 

From this derivation, we can deduce two implicit assumptions embedded in SILO's procedure.

\implicitassumption{ia:suff}{Encoder Reconstruction Quality}
The guidance signal~\eqref{eq:silo_loss} works entirely with $\mc{E}(\mby)$
instead of $\mby$. SILO justifies this by showing that the autoencoder
reconstructs measurements well 
(high PSNR of $\mc{D}(\mc{E}(\mby))$ vs $\mby$) for common image processing tasks.

\implicitassumption{ia:gauss}{Isotropic Gaussian Likelihood}
Denoting $\mu_\theta(\mbz_{0 \mid t}) = H_\theta(\mbz_{0\mid t})$,
SILO's guidance~\eqref{eq:silo_loss} takes the form
$-\nabla_{\mbz_t} \norm{\mc{E}(\mby) - \mu_\theta(\mbz_{0 \mid t})}{2}^2$.
We observe that this is, up to a positive scalar, the score of an isotropic Gaussian: 

\begin{proposition}\label{prop:silo_special}
    Suppose $p(\mc{E}(\mby) \mid \mbz_t) = \mc{N}(\mc{E}(\mby);\, \mu_\theta(\mbz_{0 \mid t}),\, \sigma^2 \mbf{I})$
    for some fixed $\sigma > 0$. Then, the score is $\nabla_{\mbz_t} \log p(\mc{E}(\mby) \mid \mbz_t)
        = \frac{1}{\sigma^2}\,
        (\nabla_{\mbz_t} \mu_\theta(\mbz_{0 \mid t}))^\top (\mc{E}(\mby) - \mu_\theta(\mbz_{0 \mid t}))$,
        recovering SILO's guidance~\eqref{eq:silo_loss} with scale $\eta = 1/(2\sigma^2)$.
\end{proposition}

SILO is therefore implicitly modeling $p(\mc{E}(\mby) \mid \mbz_t)$ as a Gaussian
with isotropic, input-independent covariance. 

\FloatBarrier
\subsection{\ours{} as a Relaxation of SILO}

We can view \ours{} as a relaxation of the two implicit assumptions made in SILO. 

Specifically, considering Implicit Assumption
\ref{ia:suff}, we notice that reconstruction quality of the autoencoder on $\mby$ 
is not the same as statistical sufficiency for posterior sampling, which is what Approximation~\ref{apprx:suff} requires.
For example, a lossy encoder can achieve high reconstruction PSNR
yet still discard information relevant to $\mbz_t$. This perspective is important for future work that may consider new encoding maps that allow
for generalization to tasks beyond image-space inverse problems, e.g., phase retrieval. 

Regarding Implicit Assumption~\ref{ia:gauss}, as we demonstrated in Proposition~\ref{prop:silo_special}, SILO corresponds to the special time and input-independent case $\Sigma_\theta \equiv \sigma^2 \mbf{I}$. As we demonstrate in Sections~\ref{sec:theory} and~\ref{sec:experiments},
the lifting of this assumption to include richer covariance structure is both provably necessary in certain settings and empirically beneficial. Further, SILO is derived from the lens that for efficiency, we can learn a forward operator directly operating in the latent space: from this perspective, there is no need for the latent forward operator to be time-dependent. In fact, SILO also experiments with a time-independent CNN architecture for $H_\theta$, but adopts a time-dependent Readout-Guidance \citep{luo2024readout} architecture in general. Our framework exposes that time-dependency is not a curious artifact of the architecture, but \emph{necessary} if we aim to approximate the likelihood $p(\mc{E}(\mby) \mid \mbz_t)$ which is a time-varying likelihood. This is further illustrated in Figure~\ref{tab:ablation_variance}, which shows that using a time-independent covariance harms performance in practice. 

\FloatBarrier
\section{Extended Related Work}
\label{app:related_work}

\paragraph{Uncertainty in Pixel-Space Diffusion Models.} We review here in more detail the rich line of work of exploiting uncertainty estimates for guidance in pixel-space diffusion models, and we contrast this with the more challenging LDM setting. Recall from Section~\ref{sec:related_work} that pixel-space diffusion models consider a diffusion process directly in pixel space. The corresponding conditional score for solving inverse problems then becomes
\begin{equation}
    \grad_{\mbx_t} \log p(\mbx_t \mid \mby) = \grad_{\mbx_t} \log p(\mbx_t) + \grad_{\mbx_t} \log p(\mby \mid \mbx_t),
\end{equation}
where the noisy likelihood score requires approximation. This distribution factorizes as
\begin{equation} \label{eq:pixel_factorization}
    p(\mby \mid \mbx_t) = \int p(\mby \mid \mbx_0) p(\mbx_0 \mid \mbx_t) \ d \mbx_0.
\end{equation}
As we know the forward model $p(\mby \mid \mbx_0)$, the main challenge is in approximating the posterior denoising distribution. Early methods such as DPS \citep{chung2022diffusion} approximate $p(\mbx_0 \mid \mbx_t)$ as a Dirac delta around the posterior mean $\E[\mbx_0 \mid \mbx_t]$, and Pi-GDM \citep{song2023pseudoinverse} extends to use a Gaussian distribution centered at the posterior mean with isotropic covariance. This was extended to richer covariance models, leveraging higher-order Tweedie's formulas as in TMPD \citep{boys2023tweedie} or approximating via diagonal matrices as in Peng \citep{peng2024improving} or FreeHunch \citep{rissanen2024free}. Certain variants also proposed to learn the posterior covariance by parameterizing the covariance by a deep network \citep{meng2021estimating}. Crucially, these more complex covariance models such as TMPD, Peng, and FreeHunch are only applicable when the forward operator $\mc{A}$ is linear, otherwise the integral in \ref{eq:pixel_factorization} is still intractable. 

This issue is precisely why these more complex posterior covariance models do not aid in modeling the noisy likelihood score for LDMs . Namely, for LDMs, Equation~\ref{eq:pixel_factorization} becomes
\begin{equation}
    p(\mby \mid \mbz_t) = \int p(\mby \mid \mbz_0) p(\mbz_0 \mid \mbz_t) \ d \mbz_0.
\end{equation}
Even with Gaussian parameterizations of $p(\mbz_0 \mid \mbz_t)$, the likelihood $p(\mby \mid \mbz_0)$ is $\mc{N}(\mc{A}(\mc{D}(\mbz_0)), \sigma_y^2 \mbf{I})$. Because the decoder is nonlinear, this integral no longer maintains any closed form. As such, prior works are mostly limited to considering approximations to the noisy likelihood evaluated at the posterior mean only. Attempts to include uncertainty have focused on using surrogate guidance terms that include regularizations on second-order estimates of uncertainty. Specifically, STSL \citep{rout2024beyond} uses a surrogate guidance that targets uncertainty by the second-order Tweedie formula, which links denoising covariance to the Jacobian of the denoiser. The norm of this Jacobian, estimated via finite difference evaluations of the denoiser in a perturbation set, is then added as a regularizer. However, these finite difference evaluations greatly increase sampling time, increasing denoiser evaluations by five to ten fold during every guidance step. Our work avoids these closed form computations and additional regularizations by directly targeting the likelihood in the latent space $p(\mc{E}(\mby) \mid \mbz_t)$, which permits a straightforward learning rule for the covariance. 

\paragraph{Uncertainty in Classification and Regression.} Our work draws on a mature line of research on heteroscedastic uncertainty estimation in supervised learning. The foundational idea of training a neural network to predict both a mean and an input-dependent variance dates back to \citet{nix1994estimating}, who proposed minimizing the Gaussian negative log-likelihood (NLL) so that the network learns to output larger variance on inherently noisy inputs. \citet{kendall2017uncertainties} popularized this approach in deep learning for computer vision, distinguishing \emph{aleatoric} uncertainty (irreducible noise inherent to the data) from \emph{epistemic} uncertainty (model uncertainty reducible with more data), and showing that modeling aleatoric uncertainty via a learned heteroscedastic variance improves both regression and classification. More recently, \citet{seitzer2022pitfalls} identified optimization pitfalls in heteroscedastic NLL training---specifically, that the network can exploit the learned variance to explain away large residuals rather than fitting the mean---and proposed $\beta$-NLL, which reweights the loss to mitigate this failure mode.

Our framework can be viewed as an adaptation of this heteroscedastic paradigm from supervised prediction to diffusion posterior sampling. In the classification/regression setting, one trains a network to output $(\mu_\theta(\mbx), \sigma^2_\theta(\mbx))$ and minimizes $\frac{1}{2}\log \sigma^2_\theta + \frac{1}{2\sigma^2_\theta}\|y - \mu_\theta\|^2$. Analogously, \ours{} trains a latent forward operator to output a mean $\mu_\theta(\mbz_t, t, \sigma_y)$ and a diagonal covariance $\Sigma_\theta(\mbz_t, t, \sigma_y)$ and minimizes the multivariate Gaussian NLL (Equation~\ref{eq:nll}). The key difference is the downstream use: rather than producing calibrated predictive intervals, the learned covariance is used to construct a Mahalanobis-weighted guidance signal that steers diffusion sampling. In this sense, \ours{} transfers the benefits of heteroscedastic uncertainty estimation---emphasizing reliable predictions and downweighting uncertain ones---from discriminative tasks to generative posterior sampling with diffusion models. We also note that in our two-stage training procedure, we first train the mean with a frozen isotropic covariance before learning the covariance, which sidesteps the optimization pitfalls identified by \citet{seitzer2022pitfalls}, as the mean network is not incentivized to exploit the variance head during its own training.

% ============================================================
\FloatBarrier
\section{Proofs from Section~\ref{sec:theory}}
\label{app:proofs}
% ============================================================

\subsection{Proof of
Proposition~\ref{prop:autoenc_optim}
(Optimal Autoencoder)}
\label{app:proof_autoenc}

\begin{proof}
We prove global optimality by direct substitution and a
lower bound argument.

First, recall the encoder and decoder are of the form:
\begin{align}
  \mc{E}(\mbx) &= \mbf{W} (\mbf{g} \Mconv \mbx) \\
  \mc{D}(\mbz) &= \mbf{h} \MconvT (\mbf{Vz})
\end{align}
Suppose the weights of these maps satisfy the conditions of the proposition, so
\begin{equation}
  (\mbf{F}_d \mbf{g})_\omega =
  \begin{cases}
  \sqrt{M}, & \omega \in \Omega_k,\\
  0, & \omega \notin \Omega_k,
  \end{cases}
  \qquad
  (\mbf{F}_d \mbf{h})_\omega =
  \begin{cases}
  \sqrt{M}, & \omega \in \Omega_k,\\
  0, & \omega \notin \Omega_k,
  \end{cases}
  \end{equation}
and 
\begin{align}
  \mbf{W} &= \mbf{F}_k\,
  \operatorname{diag}_{\omega\in\Omega_k}
  (\lambda_\omega^{-1/2})\,
  \mbf{F}_k^*, \label{eq:W_opt_app} \\[4pt]
  \mbf{V} &= \mbf{F}_k\,
  \operatorname{diag}_{\omega\in\Omega_k}
  (\lambda_\omega^{1/2})\,
  \mbf{F}_k^*. \label{eq:V_opt_app}
  \end{align}
Then, the encoder acts an ideal low-pass filter followed by
whitening of the retained Fourier coefficients, and the decoder 
is an inverse of the encoder on the subspace spanned by the first
$k$ (centered) frequencies. Thus, the composition $\mc{D} \circ \mc{E}$
is the orthogonal projector $P_{\Omega_k}$ onto the subspace
spanned by the frequencies in $\Omega_k$. Therefore the
reconstruction error is
\begin{equation} \label{eq:optimal_vae_recon}
\E\|\mc{D}(\mc{E}(\mbx_0))-\mbx_0\|_2^2
=
\sum_{\omega\in\Omega_d\setminus\Omega_k}
\lambda_\omega.
\end{equation} 
Further, because of the whitening, the pushforward  $\mc{E}_\sharp p$
is exactly a standard $k$-dimensional Gaussian, so the KL-term
in the objective is exactly $0$. 

We now show that the above setting of the encoder and decoder
is a global minimizer of \eqref{eq:vae_loss}. To show this, we will
demonstrate that \eqref{eq:optimal_vae_recon} matches a lower bound on the reconstruction error 
of any encoder/decoder pair. This fact combined with the non-negativity
of the KL divergence will prove that the above setting must be a global
minimizer of \eqref{eq:vae_loss}.

The lower bound of reconstruction error follows from the 
Eckart-Young-Mirsky theorem~\citep{eckart1936approximation}. Specifically, 
because $\mc{E}$ is a linear map with codomain 
$\R^k$ where $k < d$, the composition $\mc{D} \circ \mc{E}$ has 
rank at most $k$ and thus the reconstruction error is minimized
by retaining the $k$ largest-variance principal components of the data.
By Assumption~\ref{ass:data}, due to the power law distribution
of the data in frequency domain, the top principal components are 
precisely the lowest-magnitude centered frequencies (the set $\Omega_k$).
Since the setting of the encoder and decoder given above precisely
performs this projection, it must be a global minimizer of the reconstruction
loss and therefore also a global minimizer of \eqref{eq:vae_loss} by 
non-negativity of the KL loss.
\end{proof}

\begin{remark}[Role of the isotropic latent prior]
At the optimal autoencoder, $\mc{E}_\sharp p = \mc{N}(\mbf{0}, \mbf{I}_k)$,
so the latent marginal $p(\mbz_t)$ is standard Gaussian for all~$t$ and the
unconditional diffusion process is trivial (the score is simply $-\mbz_t$).
This is by design: the purpose of our theoretical model is not to study
unconditional generation, but to characterize the \emph{likelihood}
$p(\mc{E}(\mby) \mid \mbz_t)$ and its covariance structure,
which is the object that governs guidance.
The covariance anisotropy (Proposition~\ref{prop:sufficiency}, Part~\ref{prop:latent_lik}) arises
entirely from the interaction of the forward operator, the encoder's
whitening, and the diffusion timestep---none of which depend on the
latent prior being non-trivial.
The isotropic prior is an analytical convenience that makes the
posterior mean available in closed form via Tweedie's formula;
in practice, LDM encoders do not achieve $\mathrm{KL} = 0$
and the latent distribution retains further structure.
\end{remark}

\subsection{Proof of
Proposition~\ref{prop:sufficiency}}
\label{app:proof_sufficiency}
\label{app:proof_latent_lik}

We prove Proposition~\ref{prop:sufficiency} by establishing
two lemmas, one for each part, and then combining them.

\begin{lemma}[Sufficiency of encoded measurements]
\label{lem:sufficiency}
Under Assumptions~\ref{ass:data}--\ref{ass:ip} with the
optimal autoencoder
(Proposition~\ref{prop:autoenc_optim}),
$\mc{E}(\mby)$ is a sufficient statistic for $\mbz_t$
given $\mby$:
\begin{equation}
p(\mbz_t \mid \mby) = p(\mbz_t \mid \mc{E}(\mby)).
\end{equation}
\end{lemma}

\begin{proof}
Since $\mbA$ is a circular convolution as per
Assumption~\ref{ass:ip}, the measurement decomposes
component-wise in centered Fourier coordinates:
\begin{equation}
(\mbf{F}_d \mby)_\omega
= a_\omega (\mbf{F}_d \mbx_0)_\omega
+ \sigma_y (\mbf{F}_d \bm{\epsilon})_\omega,
\qquad \omega \in \Omega_d,
\end{equation}
where $a_\omega$ is the Fourier response of $\mbA$ at index $\omega$.
We denote $\mby^{\mathrm{lo}}$ for the Fourier coefficients
indexed by $\Omega_k$, and $\mby^{\mathrm{hi}}$ for the
remaining coefficients indexed by
$\Omega_d \setminus \Omega_k$.

Under the optimal autoencoder from
Proposition~\ref{prop:autoenc_optim}, $\mc{E}(\mby)$ is
an invertible whitening transform of
$\mby^{\mathrm{lo}}$ (since $\mbf{W}$ is invertible),
so conditioning on $\mc{E}(\mby)$ is equivalent to
conditioning on $\mby^{\mathrm{lo}}$.

The latent variable $\mbz_0 = \mc{E}(\mbx_0)$ depends
only on the coefficients of $\mbx_0$ in $\Omega_k$, and
$\mbz_t$ depends on $\mbx_0$ only through $\mbz_0$.
Under Assumption~\ref{ass:data}, the Fourier coefficients
of $\mbx_0$ are independent across frequencies, and the
measurement noise is independent across frequencies.
Therefore the high-frequency measurement
$\mby^{\mathrm{hi}}$ is conditionally independent of
$\mbz_t$ given $\mby^{\mathrm{lo}}$:
\begin{equation}
\mbz_t \perp \mby^{\mathrm{hi}}
\mid \mby^{\mathrm{lo}}.
\end{equation}
Hence
\begin{equation}
p(\mbz_t\mid \mby)
= p(\mbz_t\mid \mby^{\mathrm{lo}})
= p(\mbz_t\mid \mc{E}(\mby)),
\end{equation}
which proves sufficiency.
\end{proof}

\begin{lemma}[Gaussian latent likelihood]
\label{lem:latent_lik}
Under Assumptions~\ref{ass:data}--\ref{ass:ip} with the
optimal autoencoder
(Proposition~\ref{prop:autoenc_optim}),
$p(\mc{E}(\mby) \mid \mbz_t)$ is a Gaussian distribution
with mean $\bm{\mu}_t$ and covariance $\mbf{C}_t$ given
by:
\begin{equation}
    \bm{\mu}_t = \mbf{F}_k^*\,
    \operatorname{diag}_{\omega\in\Omega_k}
    (a_\omega)\, \mbf{F}_k \mbz_{0 \mid t}, \qquad
    \mbf{C}_t = \mbf{F}_k^*\,
    \operatorname{diag}_{\omega\in\Omega_k}
    \bigl(c_\omega(t)\bigr)\, \mbf{F}_k,
\end{equation}
where $c_\omega(t) =
(1 - \bar{\alpha}_t)|a_\omega|^2
+ \sigma_y^2/\lambda_\omega$ for
$\omega \in \Omega_k$.
\end{lemma}

\begin{proof}
We observe that the latent likelihood factorizes as
\begin{equation} \label{eq:like_marginal}
p(\mc{E}(\mby) \mid \mbz_t) = \int p(\mc{E}(\mby) \mid \mbz_0) p(\mbz_0 \mid \mbz_t) \ d\mbz_0
\end{equation}
Thus, the strategy for deriving $p(\mc{E}(\mby) \mid \mbz_t)$
is to (i)~express $\mc{E}(\mby)$ as a linear function of
$\mbz_0$ plus noise (the \emph{latent measurement model}),
and then (ii)~marginalize out $\mbz_0$ using the
posterior $p(\mbz_0 \mid \mbz_t)$.

To derive the latent measurement model, we would like to find
a commuting diagram (Figure~\ref{fig:commuting}) such that
$\mc{E}(\mbA \mbx_0) = \mbf{H}\, \mc{E}(\mbx_0) = \mbf{H}\mbz_0$.
While this motivates our learning objective in general, in our
setting, we can exploit spectral properties of the autoencoder
and the forward operator to find a closed form of $\mbf{H}$. Specifically,
we use the Noble identities from multirate signal processing \citep{vaidyanathan2006multirate}.
The Noble identities show that filtering followed by downsampling commutes with downsampling
followed by an appropriately aliased filter.
In our setting, since the encoder's ideal low-pass filter eliminates all frequencies
outside $\Omega_k$ before downsampling, and $\mbA$ acts independently on each frequency,
the two operations commute exactly: $\mc{E} \circ \mbA = \mbf{H} \circ \mc{E}$,
where $\mbf{H}$ retains only the action of $\mbA$ on the frequencies in $\Omega_k$.
Expressed mathematically,
\begin{equation}
    \mbf{H} = \mbf{F}_k^*\,
    \operatorname{diag}_{\omega\in\Omega_k}(a_\omega)\,
    \mbf{F}_k.
\end{equation}

For the noise term, $\bm{\epsilon} \sim \mc{N}(\mbf{0}, \mbf{I}_d)$
has flat spectrum, and the encoder whitens each retained
frequency by $\lambda_\omega^{-1/2}$, amplifying the
noise non-uniformly.
Combining both terms gives the latent measurement model:
\begin{equation}\label{eq:latent_meas_model}
    \mc{E}(\mby) = \mbf{H}\mbz_0 + \bm{\eta},
    \qquad \bm{\eta} \sim \mc{N}\!\left(\mbf{0},\,
    \sigma_y^2\, \mbf{F}_k^*\,
    \operatorname{diag}_{\omega\in\Omega_k}
    (\lambda_\omega^{-1})\, \mbf{F}_k\right),
    \quad \bm{\eta} \perp \mbz_0.
\end{equation}
This is a standard linear-Gaussian observation model
in the latent space, with a diagonal operator and
anisotropic noise whose amplification
$\sigma_y^2/\lambda_\omega$ grows for higher frequencies.

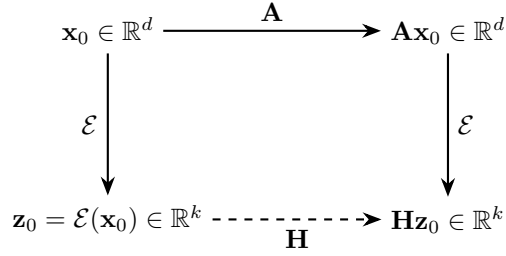
\begin{figure}[t!]
\centering
\begin{tikzpicture}[
    every node/.style={font=\normalsize},
    arr/.style={-{Stealth[length=2.5mm]}, thick}
]
\node (x) at (0,0) {$\mbx_0 \in \R^d$};
\node (Ax) at (4.5,0) {$\mbA\mbx_0 \in \R^d$};
\node (z) at (0,-2.5) {$\mbz_0 = \mc{E}(\mbx_0) \in \R^k$};
\node (Hz) at (4.5,-2.5) {$\mbf{H}\mbz_0 \in \R^k$};

\draw[arr] (x) -- node[above] {$\mbA$} (Ax);
\draw[arr] (x) -- node[left] {$\mc{E}$} (z);
\draw[arr] (Ax) -- node[right] {$\mc{E}$} (Hz);
\draw[arr, dashed] (z) -- node[below] {$\mbf{H}$} (Hz);
\end{tikzpicture}
\caption{Commuting diagram for the latent forward operator.
The Noble identity guarantees that encoding after degradation
($\mc{E} \circ \mbA$, right then down) equals latent degradation
after encoding ($\mbf{H} \circ \mc{E}$, down then right),
so the diagram commutes:
$\mc{E}(\mbA\mbx_0) = \mbf{H}\,\mc{E}(\mbx_0) = \mbf{H}\mbz_0$.}
\label{fig:commuting}
\end{figure}

Next, we can marginalize using \eqref{eq:like_marginal}.
Under the
variance-preserving noising process,
the posterior is
$p(\mbz_0 \mid \mbz_t)
= \mc{N}(\hat{\mbz}_{0\mid t},\,
(1-\bar\alpha_t)\,\mbf{I}_k)$,
where
$\hat{\mbz}_{0\mid t}
\coloneqq \E[\mbz_0 \mid \mbz_t]
= \sqrt{\bar\alpha_t}\,\mbz_t$.
The last equality follows from Tweedie's formula:
since $\mbz_0 \sim \mc{N}(\mbf{0}, \mbf{I}_k)$,
$p(\mbz_t)$ is also standard Gaussian and
$\grad_{\mbz_t}\!\log p(\mbz_t) = -\mbz_t$, giving
$\E[\mbz_0 \mid \mbz_t]
= \frac{1}{\sqrt{\bar\alpha_t}}
\bigl(\mbz_t + (1-\bar\alpha_t)(-\mbz_t)\bigr)
= \sqrt{\bar\alpha_t}\,\mbz_t$. Conditioning on $\mbz_t$, the latent measurement
model~\eqref{eq:latent_meas_model} becomes a linear
function of the Gaussian $\mbz_0 \mid \mbz_t$ plus
independent Gaussian noise $\bm{\eta}$.
By standard Gaussian marginalization,
\begin{equation}
    p(\mc{E}(\mby) \mid \mbz_t)
    = \mc{N}\!\left(\bm{\mu}_t,\;
    \mbf{C}_t\right),
\end{equation}
where the mean is
$\bm{\mu}_t \coloneqq \mbf{H}\hat{\mbz}_{0\mid t}$
and the covariance collects two sources of uncertainty:
\begin{equation}
\mbf{C}_t =
\underbrace{(1-\bar\alpha_t)\,\mbf{H}\mbf{H}^*}_{%
\text{diffusion uncertainty through } \mbA}
\;+\;
\underbrace{\sigma_y^2\, \mbf{F}_k^*\,
\operatorname{diag}_{\omega\in\Omega_k}
(\lambda_\omega^{-1})\, \mbf{F}_k}_{%
\text{measurement noise amplified by whitening}}.
\end{equation}
Since $\mbf{H}$ is diagonal in the latent Fourier basis,
the covariance eigenvalues are
\begin{equation}
    c_\omega(t) =
    (1 - \bar\alpha_t)\, |a_\omega|^2
    + \frac{\sigma_y^2}{\lambda_\omega},
    \qquad \omega \in \Omega_k.
\end{equation}
Both terms vary across frequencies, making the covariance
inherently anisotropic.
\end{proof}

\begin{proof}[Proof of Proposition~\ref{prop:sufficiency}]
Part~1 follows directly from Lemma~\ref{lem:sufficiency},
and Part~\ref{prop:latent_lik} follows directly from
Lemma~\ref{lem:latent_lik}.
\end{proof}

\subsection{Proof of
Theorem~\ref{thm:scalar_approx}
(Isotropic vs.\ Anisotropic Covariance
Approximation Error)}
\label{app:proof_scalar_approx}

\begin{proof}

\textbf{Part 1: Irreducible error of isotropic covariance.}\;
The isotropic approximation replaces the true covariance
$\mbf{C}_t$ with $\eta^2\,\mbf{I}_k$ for a scalar
$\eta > 0$, while keeping the mean $\bm{\mu}_t$
unchanged. We want to show that the KL divergence
between the true and approximate likelihoods is
lower bounded by a strictly positive quantity for
any choice of $\eta$.

Since both the true and approximate likelihoods are
Gaussians with the same mean, the KL divergence
depends only on the covariances. Recall that for
two Gaussians with shared mean,
\begin{equation}
\KL{\mc{N}(\bm{\mu}, \bm{\Sigma}_1)}
    {\mc{N}(\bm{\mu}, \bm{\Sigma}_2)}
= \frac{1}{2}\bigl[
    \tr(\bm{\Sigma}_2^{-1}\bm{\Sigma}_1)
    + \log\det\bm{\Sigma}_2
    - \log\det\bm{\Sigma}_1
    - k
\bigr].
\end{equation}
We set $\bm{\Sigma}_1 = \mbf{C}_t$ and
$\bm{\Sigma}_2 = \eta^2\,\mbf{I}_k$.
Since both are diagonalized by the Fourier basis
$\mbf{F}_k$, we can evaluate each term using the
eigenvalues $\{c_\omega(t)\}_{\omega \in \Omega_k}$
directly:
\begin{align}
\tr(\bm{\Sigma}_2^{-1}\bm{\Sigma}_1)
&= \frac{1}{\eta^2}
   \sum_{\omega\in\Omega_k} c_\omega, \\
\log\det\bm{\Sigma}_2
&= k\log\eta^2, \\
\log\det\bm{\Sigma}_1
&= \sum_{\omega\in\Omega_k}\log c_\omega.
\end{align}
Substituting, the KL divergence is
\begin{equation}\label{eq:kl_scalar}
    \KL{p(\mc{E}(\mby) \mid \mbz_t)}
        {\mc{N}(\bm{\mu}_t,\,
        \eta^2\,\mbf{I}_k)}
    = \frac{1}{2}\!\left[
        \frac{1}{\eta^2}
        \sum_{\omega\in\Omega_k} c_\omega
        + k\log\eta^2
        - \sum_{\omega\in\Omega_k}\log c_\omega
        - k
    \right].
\end{equation}
We now minimize over $\eta^2$ to find the tightest
possible lower bound.
Differentiating~\eqref{eq:kl_scalar} with respect to
$\eta^2$ and setting the derivative to zero:
\begin{align}
\frac{\partial}{\partial \eta^2}
\left[
    \frac{1}{\eta^2}
    \sum_{\omega} c_\omega + k\log\eta^2
\right]
&= -\frac{1}{\eta^4}
   \sum_{\omega} c_\omega + \frac{k}{\eta^2}
= 0 \\[4pt]
\Longrightarrow\quad
\eta^{*2}
&= \frac{1}{k}
   \sum_{\omega\in\Omega_k} c_\omega(t)
= \mathrm{AM}\!\bigl(
   \{c_\omega(t)\}_{\omega\in\Omega_k}\bigr).
\end{align}
The second derivative
$\frac{2}{\eta^6}\sum_\omega c_\omega
- \frac{k}{\eta^4}$
is positive at $\eta^{*2}$, confirming this is a
minimum. Substituting $\eta^{*2}$ back
into~\eqref{eq:kl_scalar}, the trace term becomes
$\frac{1}{\eta^{*2}}\sum_\omega c_\omega = k$,
which cancels the $-k$ term. The remaining expression
is
\begin{align}
\frac{k}{2}\log\mathrm{AM}(
\{c_\omega\}_{\omega\in\Omega_k})
- \frac{1}{2}
\sum_{\omega\in\Omega_k}\log c_\omega
&= \frac{k}{2}\log\mathrm{AM}(
   \{c_\omega\}_{\omega\in\Omega_k})
- \frac{k}{2}\log\mathrm{GM}(
   \{c_\omega\}_{\omega\in\Omega_k}) \\
&= \frac{k}{2}\log
   \frac{\mathrm{AM}(
   \{c_\omega\}_{\omega\in\Omega_k})}
   {\mathrm{GM}(
   \{c_\omega\}_{\omega\in\Omega_k})},
\end{align}
where we used the identity
$\sum_\omega \log c_\omega
= k \log \mathrm{GM}(
\{c_\omega\}_{\omega\in\Omega_k})$.
By the AM--GM inequality,
$\mathrm{AM} \geq \mathrm{GM}$
with equality if and only if all $c_\omega$ are equal,
establishing the bound~\eqref{eq:kl_amgm}.

It remains to show strict positivity under
Assumptions~\ref{ass:data}--\ref{ass:ip}.
The bound~\eqref{eq:kl_amgm} equals zero if and only if
all eigenvalues $c_\omega(t)$ are identical.
We show this cannot happen.

At $t = 0$ ($\bar\alpha_0 = 1$), the diffusion term
vanishes and $c_\omega(0) = \sigma_y^2/\lambda_\omega$.
Since $\lambda_\omega = C(1+|\omega|)^{-\alpha}$ with
$\alpha > 1$, this equals
$(\sigma_y^2/C)(1+|\omega|)^\alpha$, which is strictly
increasing in $|\omega|$ and therefore non-constant
(using $\sigma_y > 0$ from Assumption~\ref{ass:ip}).

For $t > 0$ ($\bar\alpha_t < 1$), the eigenvalues are
$c_\omega(t)
= (1 - \bar\alpha_t)\,|a_\omega|^2
+ \sigma_y^2/\lambda_\omega$.
For these to be constant across $\omega$, we need
$|a_\omega|^2 = (K - \sigma_y^2/\lambda_\omega)/(1-\bar\alpha_t)$
for some constant $K$. Since $|a_\omega|^2$ is fixed while
$\bar\alpha_t$ varies with $t$, this can hold for at most one
$t \in (0, T)$, so under $t \sim \mc{U}[0, T]$, this is a
measure-zero event.

Therefore the
$\{c_\omega(t)\}_{\omega\in\Omega_k}$ are non-constant
with probability 1 when $t \sim \mc{U}[0,T]$, and the KL
divergence~\eqref{eq:kl_amgm} is strictly positive.

\textbf{Part 2: Convergence of diagonal covariance model.}\;
In the Fourier basis, the true likelihood decomposes
into $k$ independent 1D Gaussians: the $\omega$-th
component has mean
$a_\omega\,(\mbf{F}_k\mbz_{0\mid t})_\omega$
and variance $c_\omega(t)$.
A diagonal Gaussian model with linear mean and free
diagonal covariance matches this structure exactly, so
the model is well-specified.

For each frequency $\omega$, the MLE estimates the
coefficient $a_\omega$ via ordinary least squares and the
variance $c_\omega(t)$ as the sample variance of
residuals. Standard results for Gaussian MLE on a
well-specified model \citep{lehmann1998theory} give, for
each component,
$\E[\mathrm{KL}_\omega] = O(1/n)$,
where the implicit constant depends on the true
parameters but is bounded for each fixed $t$.
Summing over $k$ independent components yields
$\E\bigl[\KL{p(\mc{E}(\mby) \mid \mbz_t)}
{\hat{p}(\mc{E}(\mby) \mid \mbz_t)}\bigr]
= O(k/n)$.
\end{proof}

% ============================================================
\FloatBarrier
\section{Discussion of Theoretical Analysis}
\label{app:theory_extended}
% ============================================================

\FloatBarrier
\subsection{Regimes Where Isotropic Approximation Error
Is Large}
\label{app:theory_analysis}

The KL lower bound~\eqref{eq:kl_amgm} equals
$\frac{k}{2}\log(\mathrm{AM}/\mathrm{GM})$, which
measures the \emph{multiplicative spread} of the
eigenvalues
$\{c_\omega(t)\}_{\omega\in\Omega_k}$. The discussion
below identifies regimes in which
Theorem~\ref{thm:scalar_approx} predicts this quantity
to be $\Omega(1)$ or larger.

To understand the scaling of the lower bound, recall that
$c_\omega(t) = (1 - \bar\alpha_t)\,|a_\omega|^2
+ \sigma_y^2/\lambda_\omega$.
Since $\lambda_\omega \propto (1+|\omega|)^{-\alpha}$,
the second term scales as
$\sigma_y^2/\lambda_\omega
\propto (1+|\omega|)^{\alpha}$.
Let $\omega_{\max} \in \Omega_k$ denote a retained
frequency with largest magnitude,
$|\omega_{\max}| = (k-1)/2$.
The ratio between the largest and smallest eigenvalues
satisfies
\begin{equation}\label{eq:dyn_range}
    \frac{c_{\omega_{\max}}(t)}{c_0(t)}
    \;\geq\;
    \frac{\sigma_y^2 / \lambda_{\omega_{\max}}}
    {\sigma_y^2 / \lambda_0
    + (1-\bar\alpha_t)\,\|a\|_\infty^2}
    \;=\;
    \frac{(1+|\omega_{\max}|)^\alpha}
    {1 + (1-\bar\alpha_t)\,\|a\|_\infty^2\,
    \lambda_0/\sigma_y^2},
\end{equation}
where $\|a\|_\infty = \max_\omega |a_\omega|$. This
tells us that the eigenvalue ratio grows exponentially in the spectral decay~$\alpha$. Since the lower bound is the logarithm of this quantity, the lower bound grows linear in the spectral decay. 

\paragraph{Near the data end of the reverse process
($t \to 0$, $\bar\alpha_t \approx 1$).}
The diffusion noise contribution
$(1 - \bar\alpha_t)\,|a_\omega|^2$ is small, so the
eigenvalue spread is dominated by
$\sigma_y^2/\lambda_\omega$ and the isotropic error is
largest. As the reverse process moves towards the data
($t$ decreases, $\bar\alpha_t \to 1$), the diffusion term
$(1 - \bar\alpha_t)\,|a_\omega|^2$ shrinks, leaving the
eigenvalue spread increasingly dominated by the
measurement noise term $\sigma_y^2/\lambda_\omega$, which
varies strongly across frequencies. This \emph{increases}
the isotropic penalty near the data end, consistent with
Fig.~\ref{fig:kl_bound}.

\paragraph{Beyond convolutional operators.}
Our theoretical analysis focuses on deconvolution, where
the convolutional structure of the forward operator and
autoencoder enables exact characterization. In practice,
native-latent guidance also applies to non-convolutional
inverse problems such as inpainting and JPEG
restoration, where the sufficiency and likelihood
structure may not admit closed-form expressions. We
verify empirically in Section~\ref{sec:experiments} that
the anisotropic covariance structure improves
performance on these tasks as well; extending the
theoretical analysis to non-convolutional operators is
an interesting direction for future work.

\FloatBarrier
\subsection{Additional Visualizations}
\label{app:theory_viz}

We provide additional visualizations that complement the
analysis of the covariance eigenvalues $c_\omega(t)$
from Proposition~\ref{prop:sufficiency} and the isotropic
KL lower bound from
Theorem~\ref{thm:scalar_approx}. All plots use data
dimension $d = 256$, latent dimension $k = 64$,
$\sigma_y = 0.05$, $\alpha = 2.5$ (unless varied), and
the Gaussian blur operator from the main text.

\begin{figure}[ht]
    \centering
    \includegraphics[width=0.48\linewidth]
    {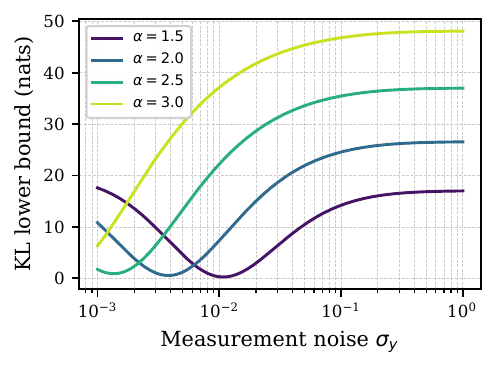}
    \caption{KL lower bound~\eqref{eq:kl_amgm} vs.\
    measurement noise $\sigma_y$ at $t = 0.05$ for
    different spectral decay rates~$\alpha$. The
    isotropic penalty depends non-trivially on
    $\sigma_y$: at low noise the bound is set by the
    operator's frequency response, while at high noise
    the encoder's whitening
    ($\sigma_y^2/\lambda_\omega$) dominates.}
    \label{fig:kl_vs_sigma}
\end{figure}

\paragraph{Effect of measurement noise $\sigma_y$.}
Figure~\ref{fig:kl_vs_sigma} shows the KL lower bound
as a function of $\sigma_y$ at a fixed near-data
timestep ($t = 0.05$). Recall that
$c_\omega(t) = (1 - \bar\alpha_t)\,|a_\omega|^2
+ \sigma_y^2/\lambda_\omega$.
At very low $\sigma_y$, the noise-amplification term
$\sigma_y^2/\lambda_\omega$ is negligible and the
eigenvalue spread is determined entirely by the
frequency response $|a_\omega|^2$ of the forward
operator. For mild spectral decay ($\alpha = 1.5$), the
blur's frequency response is sufficiently non-flat to
produce a sizeable KL even at low noise. As $\sigma_y$
increases, the noise-amplification term dominates and
the eigenvalue spread grows as $(1+|\omega|)^\alpha$,
so the bound increases monotonically for
larger~$\alpha$. For intermediate $\alpha$ values, the
two regimes create a non-monotonic profile: the KL first
dips as the diffusion-noise term and the
measurement-noise term partially balance, then rises as
the noise amplification takes over. This suggests that
the optimal guidance strategy depends on the noise level
in a non-trivial way that a scalar $\eta$ cannot
capture.

\begin{figure}[ht]
    \centering
    \includegraphics[width=0.48\linewidth]
    {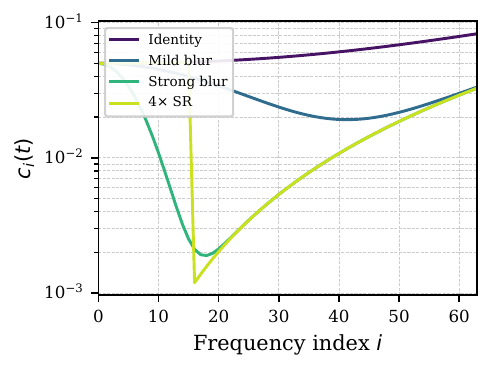}
    \caption{Covariance eigenvalues $c_\omega(t)$ at
    $t = 0.05$ for different forward operators
    ($\alpha = 2.5$, $\sigma_y = 0.001$).
    Eigenvalues are plotted against nonnegative
    frequency magnitude $|\omega|$. Each operator
    produces a distinct spectral profile. The identity
    is nearly flat; blurs suppress high frequencies
    progressively; $4\times$ SR introduces a sharp
    cutoff. An isotropic covariance cannot adapt to
    these task-dependent structures.}
    \label{fig:cov_operators}
\end{figure}

\paragraph{Dependence on the forward operator.}
Figure~\ref{fig:cov_operators} compares the covariance
spectrum $c_\omega(t)$ at a fixed timestep for four
different forward operators: the identity
($|a_\omega| = 1$ for all $\omega$), a mild Gaussian
blur (bandwidth $= 30$), a strong Gaussian blur
(bandwidth $= 8$), and $4\times$ super-resolution
($|a_\omega| = 1$ for $|\omega| < k/4$,
$|a_\omega| = 0$ otherwise). To isolate the effect of
the operator, we use a low measurement noise
$\sigma_y = 0.001$ so that the diffusion term
$(1 - \bar\alpha_t) |a_\omega|^2$ shapes the spectrum
rather than the noise floor
$\sigma_y^2/\lambda_\omega$.

The identity operator produces a nearly flat covariance
spectrum, since $|a_\omega| = 1$ for all $\omega$ and
the only variation comes from the small noise term. As
the blur strengthens, the covariance dips at frequencies
where $|a_\omega| \approx 0$: the forward operator
destroys these components, so the only remaining
uncertainty is the noise floor. Super-resolution
produces the most dramatic structure: a sharp cliff at
$|\omega| = k/4$ where the operator's frequency
response drops to zero. For $|\omega| < k/4$, the
covariance is dominated by diffusion uncertainty; for
$|\omega| \geq k/4$, it drops to the noise floor.

This visualization illustrates that different inverse
problems induce qualitatively different covariance
landscapes. An isotropic approximation applies the same
precision to all coordinates regardless of these
structural differences, while the learned 
covariance in our framework adapts to each task's
spectral profile.

%\subsection{Comparison to~\citep{rout2023solving}}
%\label{app:comparison_rout}

% Rout et al.~\citep{rout2023solving} also study a
% theoretical setting for posterior sampling with latent
% diffusion models. Our setting has several key
% differences. First, while Rout et al.\ assume perfect
% unique recoverability of the inverse problem, we study
% general deconvolution tasks, which are typically
% extremely ill-posed in practice and do not have unique
% solutions. Second, we exploit specific spectral
% properties of our data distribution in the Fourier
% basis, which gives rise to the likelihood covariance
% structure central to our main result
% (Proposition~\ref{prop:sufficiency}, Part~\ref{prop:latent_lik}, and
% Theorem~\ref{thm:scalar_approx}). This covariance
% structure is absent from the analysis
% of~\citep{rout2023solving}, which does not study the role
% of anisotropic covariance in guidance.

\FloatBarrier
\section{Experimental Details}
\label{app:exp_details}

\paragraph{Datasets, resolution, and prompts.}
Both training and evaluation use $512 \times 512$
images. On FFHQ~\citep{Karras2019ASG} (CC BY-NC-SA 4.0) we hold out 1{,}000 images for
evaluation and use the remainder for training; on
ImageNet~\citep{deng2009imagenet} (distributed under ImageNet Terms of Access for non-commercial research and educational purposes) we train on the standard train split and
evaluate on 1{,}000 randomly selected validation images.
Following SILO~\citep{raphaeli2025silo}, the FFHQ runs use the fixed
text prompt ``\emph{A high quality photo of a face}''
during sampling, whereas the ImageNet runs use an empty prompt. 
The negative prompt is empty for all runs. We do not use 
Classifier-Free Guidance (CFG) \citep{ho2022classifier}
for any of the sampling results.

\paragraph{Inverse Problem Task Descriptions.} For Gaussian deblurring, the forward operator uses a kernel of size $61 \times 61$ with standard deviation $3.0$. For super-resolution task, the forward operator uses downsampling by bicubic interpolation with antialiasing. For inpainting, we mask out the center $256\times256$ box; note this is a particularly challenging inpainting task, especially for face datasets where this box usually occludes most of the face. Thus, we cannot expect exact reconstruction and perceptual quality of reconstructions is a more faithful metric. For JPEG decompression, the forward operator compresses by a quality factor of $10$.

\FloatBarrier
\subsection{Training Details}

To foster reproducibility, we detail all architectural and 
training details below. We did not attempt to optimize 
these training hyperparameters or architectures.

\paragraph{Mean head architecture and UNet feature
aggregation.}
The mean head and variance head share a common backbone
of aggregated UNet features, following the
Readout-Guidance design~\citep{luo2024readout}.
We attach hooks to all twelve ResNet blocks in the four
upsampling stages of the SD-1.5 UNet (the up-block channel
widths are $[1280, 1280, 640, 320]$ with three ResNet
blocks each), giving a per-block feature with channel
dimensions
$[1280,\!1280,\!1280,\!1280,\!1280,\!1280,
640,\!640,\!640,\!320,\!320,\!320]$.
Each block's feature map is bilinearly resampled to
$64\times64$ (matching the latent spatial resolution),
concatenated with the broadcast sigma embedding (see
below), and passed through a per-block bottleneck
projecting to $384$ channels (a $1{\times}1{\to}3{\times}3{\to}1{\times}1$
ResNet block with timestep conditioning); the twelve
projected features are then summed with learned softmax
mixing weights $w_\ell$ (one per block, also trained).
The resulting $384$-channel aggregated feature
$\mathbf{f}_{\mathrm{agg}} \in
\mathbb{R}^{384 \times 64 \times 64}$ is concatenated
with the $1$-channel sigma embedding map and fed into the
mean head, whose architecture is
$\mathrm{Conv}(385 \!\to\! 128, 3\!\times\!3) \to
\mathrm{SiLU} \to
\mathrm{Conv}(128 \!\to\! 32, 3\!\times\!3) \to
\mathrm{SiLU} \to
\mathrm{Conv}(32 \!\to\! 4, 1\!\times\!1)$.
The output is the per-pixel mean
$\mu_\theta \in \mathbb{R}^{4 \times 64 \times 64}$;
no output activation is applied. The total trainable
parameter count is $14{,}227{,}540$ for the mean head. 
The variance head $h_\phi$ adds
another $480{,}676$ parameters and is detailed in
Table~\ref{tab:logvar_head}; it consumes the same
pre-head feature
$[\mathbf{f}_{\mathrm{agg}} \,\Vert\, \mathbf{e}_\sigma]$.
The raw output of the variance head 
is passed through a positivity-preserving transform
to yield a numerically stable variance:
\begin{align}
\sigma^2 &= \mathrm{softplus}(\mathbf{r})
    + \varepsilon_{\min}, \\
\log \sigma^2 &= \mathrm{clip}\bigl(\log \sigma^2,
    \; \ell_{\min},\; \ell_{\max}\bigr), \\
\sigma^2 &\leftarrow \exp(\log \sigma^2),
\end{align}
with variance floor $\varepsilon_{\min}=10^{-4}$ and
clipping range
$[\ell_{\min},\ell_{\max}]=[-6,4]$.
We further initialize the final convolutional
layer with zero weights. The bias is chosen such that 
the variance is isotropic unit variance
at the beginning of training after passing through the softplus activation. For our two-stage training
objective, we train using the mean-fitting objective
for 75000 steps and then train the variance-fitting
objective for 10000 steps. 

\begin{table}[h]
    \centering
    \caption{Architecture of the log-variance head
    $h_\phi$. Input is the pre-head feature
    $\mathbf{f} \in \mathbb{R}^{B \times 385
    \times 64 \times 64}$ (aggregated UNet features
    $\mathbf{f}_{\mathrm{agg}} \in \mathbb{R}^{384
    \times 64 \times 64}$ concatenated with the sigma
    embedding map $\mathbf{e}_\sigma \in
    \mathbb{R}^{1 \times 64 \times 64}$). Output is
    per-pixel $\log \sigma^2 \in \mathbb{R}^{B \times
    4 \times 64 \times 64}$.}
    \label{tab:logvar_head}
    \begin{tabular}{l l c c}
    \toprule
    \textbf{Layer} & \textbf{Operation}
    & \textbf{Output shape}
    & \textbf{\#\,Params} \\
    \midrule
    Input  & $\mathbf{f} =
    [\mathbf{f}_{\mathrm{agg}}
    \,\Vert\, \mathbf{e}_\sigma]$
    & $385 \times 64 \times 64$ & --- \\
    Conv-1 & $\mathrm{Conv2d}(385 \!\to\! 128,\,
    3\!\times\!3,\, \text{pad}=1)$
    & $128 \times 64 \times 64$ & $443{,}648$ \\
           & $\mathrm{SiLU}$
           & $128 \times 64 \times 64$ & --- \\
    Conv-2 & $\mathrm{Conv2d}(128 \!\to\! 32,\,
    3\!\times\!3,\, \text{pad}=1)$
    & $32 \times 64 \times 64$  & $36{,}896$ \\
           & $\mathrm{SiLU}$
           & $32 \times 64 \times 64$  & --- \\
    Conv-3 & $\mathrm{Conv2d}(32 \!\to\! 4,\,
    1\!\times\!1)$
    & $4 \times 64 \times 64$   & $132$ \\
    \midrule
    \multicolumn{3}{r}{\textbf{Total}}
    & $\mathbf{480{,}676}$ \\
    \bottomrule
    \end{tabular}
\end{table}

\paragraph{DCT-Domain Training.} To train the DCT-domain variant, we reuse the spatial-diagonal architecture above unchanged, but interpret the head's output as per-bin diagonal log-variance in the DCT domain. The variance head, however, is a small-kernel translation-equivariant CNN, whose smoothing prior makes sense for spatial-domain inputs since adjacent input positions are spatial neighbors. But the head's output positions index DCT frequency bins, not pixels, and the input features remain spatial; this mismatch causes the conv to incorrectly smooth across DCT-bin neighbors that have no reason to share variance (e.g., the DC bin vs. its neighbors). We make a simple adjustment: apply a 2D DCT to the variance head's input feature map before the first conv. The head then operates on DCT-domain features, with input-position $(i,j)$ corresponding to DCT bin $(i,j)$, so the conv's local-smoothing prior is applied across neighboring frequencies, a meaningful prior in DCT space, and the output's per-bin semantics match the input's per-bin semantics.

\paragraph{Variance regularization.}
We add a small quadratic penalty
$\lambda_{\mathrm{var}} \cdot
\E_{\mbz_t, t, \sigma_y}[\,\| \log \Sigma_\theta \|_2^2\,]$
to the training loss, with
$\lambda_{\mathrm{var}} = 10^{-4}$. This penalty
discourages $\log \Sigma_\theta$ from drifting toward
the clipping endpoints $\ell_{\min}$ or $\ell_{\max}$
on coordinates that receive little gradient signal,
and it has negligible effect on calibrated bins where
$|\log \Sigma_\theta| \ll 1$.

\paragraph{Measurement noise embedding.}
The measurement-noise level $\sigma_y$ is encoded by
re-using the LDM's sinusoidal timestep embedding: we map
$\sigma_y$ to a pseudo-timestep
$\tilde{t} = (\sigma_y / 0.1)\cdot 999$ and pass
$\tilde{t}$ through the SD-1.5 timestep MLP, producing a
$1280$-d vector. A learned linear layer projects this
vector to $64{\times}64 = 4096$ dimensions, which we
reshape to a single-channel spatial map
$\mathbf{e}_\sigma \in \mathbb{R}^{1\times 64\times 64}$
and apply LayerNorm. This map is broadcast-concatenated
to every block's feature before bottlenecking, so both
the aggregation backbone and the heads are conditioned
on $\sigma_y$.

\paragraph{Timestep sampling during training.}
At every training step, we sample the diffusion
timestep uniformly from
$t \sim \mathrm{Uniform}\{1, 2, \ldots, T-1\}$ with
$T = 1000$ (matching the DDPM noise schedule of the base
LDM), and form $\mathbf{z}_t$ via the variance-preserving
forward process. We use a single random timestep per
sample rather than averaging over $t$.

\paragraph{Measurement-noise sampling during training.}
For every training step we sample
$\sigma_y \sim \mathrm{Uniform}[0, 0.1]$ and apply
$\mathbf{y} = \mathcal{A}(\mathbf{x}) +
\sigma_y\,\bm{\epsilon}_y$ with
$\bm{\epsilon}_y \sim \mathcal{N}(\mathbf{0}, \mathbf{I})$.
The same sampled $\sigma_y$ is fed to the network via
the sigma embedding above. 

\paragraph{Optimizer and hyperparameters.}
Both stages use AdamW with the PyTorch defaults
($\beta_1{=}0.9$, $\beta_2{=}0.999$,
$\varepsilon{=}10^{-8}$), a constant learning rate of
$2\times 10^{-4}$, and weight decay $0$. Training is in
fp32 with no mixed precision. We use a per-GPU batch
size of $8$ (no gradient accumulation) and clip the
global parameter gradient norm to $1.0$ before each
optimizer step. 

\paragraph{Hardware and timing.}
All training, sampling, and reported timings were run on
NVIDIA RTX A5000 (24\,GB) GPUs. For FFHQ dataset, we utilize pretrained Stage 1 
checkpoints from the SILO codebase. For ImageNet, we pretrain
Stage 1, which takes roughly 28 GPU hours for an A5000 GPU. Stage 2
training takes roughly 4 GPU hours on the same GPU. 

\FloatBarrier
\subsection{Inference Details and Baselines}

\paragraph{Inference step size and guidance clipping.}
We follow SILO~\citep{raphaeli2025silo}
and normalize the guidance objective by the $\ell_2$ norm of the residual $\|\mbf{r}\|_2$, treated as a constant (i.e., detached from the computation graph so that gradients do not flow through the normalization).
Following SILO, we additionally
clip the per-coordinate guidance gradient to
$[-1, 1]$ for stability against
isolated low-variance outliers.

\paragraph{DCT-domain inference details.} For the DCT
variant of \ours{} the variance head predicts a diagonal
covariance $\Sigma_\theta = \diag(v)$ in the 2D DCT
basis, so guidance applies a per-frequency precision
weighting $\hat{\mathbf{r}} \mapsto \hat{\mathbf{r}}/v$
on the DCT residual $\hat{\mathbf{r}} = \mathrm{DCT}(\mc{E}(\mby) - \mu_\theta)$
before mapping back to latent space via the inverse DCT.
The remainder of the inference algorithm is identical
to the spatial-domain variant.

\paragraph{Hyperparameters.} To ensure fairness, we first find the optimal hyperparameters for the SILO baselines via cross-validation and then use the same hyperparameters for our spatial covariance baseline. This directly measures the impact of varying covariance parameterizations while keeping all else fixed. For our DCT covariance parameterization, the scales are tuned separately as the covariance has different magnitudes than in spatial domain and the optimal scales for SILO are not the optimal scales for \ours{}-DCT. We now summarize these hyperparameters here.
On FFHQ, the guidance scale is set to 2.0 for Gaussian blur, 3.0 for
super-resolution (4$\times$ and 8$\times$), 2.5 for
inpainting, and 1.0 for JPEG decompression. For the DCT covariance variants, these become guidance scale 1.0 for Gaussian blur, 3.0 and 2.0 for super-resolution 4$\times$ and 8$\times$ respectively, 3.0 for JPEG decompression, and 1.5 for inpainting. For ImageNet, the spatial covariance variant uses 1.0 for all tasks. 
For the DCT variant, we use scale 0.5 for deblurring and 0.7 for super-resolution tasks.  See Appendix~\ref{app:further_ablations} for details on this hyperparameter search. 

\paragraph{Baselines. } For the LDPS \citep{rout2023solving} and PSLD \citep{rout2023solving} baselines, we use the code from \url{https://github.com/LituRout/PSLD} (CC-BY-4.0). We use the default hyperparameters for all tasks, given in their codebase. The only task which they do not evaluate on is JPEG decompression, for which we use the same hyperparameters as the Gaussian deblurring task. As the STSL \citep{rout2024beyond} baseline does not provide code, we re-implement within the PSLD codebase, utilizing the same hyperparameters from their paper for all tasks. For the SILO \citep{raphaeli2025silo}, we use the code from \url{https://github.com/ronraphaeli/SILO} (CC-BY-4.0). 

\FloatBarrier
\subsection{Full Training and Inference Algorithms}
\label{app:algs}
Algorithms~\ref{alg:training}
and~\ref{alg:inference} summarize the full procedures.

\begin{algorithm}[h]
\caption{\ours{} two-stage training: Gaussian NLL on encoded measurements}
\label{alg:training}
\begin{algorithmic}[1]
\Require Pretrained LDM (encoder $\mc{E}$, denoiser
$\epsilon_\phi$, scheduler $\{\bar\alpha_t\}$);
mean head $\mu_\theta$; variance head $h_\phi$;
forward model $\mc{A}$; training set
$\{\mbx^{(i)}\}$; optimizer; noise range
$[\sigma_{\min}, \sigma_{\max}]$; regularization
weight $\lambda_{\mathrm{var}}$; gradient-norm clip
$g_{\max}$; stage-1 steps $S_1 = 75{,}000$;
stage-2 steps $S_2 = 10{,}000$.
\Statex \textbf{Stage 1: train $\mu_\theta$ with
$\Sigma_\theta \!=\! \mbf{I}$ (equivalent to SILO).}
\For{$s = 1, \ldots, S_1$}
  \State Sample $\mbx, \sigma_y, t$;\quad
  $\mby \gets \mc{A}(\mbx) + \sigma_y\, \bm{\epsilon}_y$;\quad
  $\mbz_0 \gets \mc{E}(\mbx)$;\quad $\mbf{w} \gets \mc{E}(\mby)$
  \State $\mbz_t \gets \sqrt{\bar\alpha_t}\, \mbz_0 +
  \sqrt{1 - \bar\alpha_t}\, \bm{\epsilon}$
  \State $\mbf{f} \gets$ aggregated UNet features at
  $(\mbz_t, t, \sigma_y)$;\quad
  $\mu_\theta \gets$ mean head$(\mbf{f})$
  \State $\ell_\theta^{(1)} \gets \tfrac12
  \| \mbf{w} - \mu_\theta \|_2^2$
  \Comment{Equation~\eqref{eq:nll} with
  $\Sigma_\theta\!=\!\mbf{I}$}
  \State Backpropagate $\ell_\theta^{(1)}$ through
  $\mu_\theta$ and the backbone; clip global grad
  norm to $g_{\max}$; optimizer step.
\EndFor
\Statex \textbf{Stage 2: freeze $\mu_\theta$ and the
backbone; train only $h_\phi$.}
\For{$s = 1, \ldots, S_2$}
  \State Sample $\mbx, \sigma_y, t$;\quad
  $\mby \gets \mc{A}(\mbx) + \sigma_y\, \bm{\epsilon}_y$;\quad
  $\mbz_0 \gets \mc{E}(\mbx)$;\quad $\mbf{w} \gets \mc{E}(\mby)$
  \State $\mbz_t \gets \sqrt{\bar\alpha_t}\, \mbz_0 +
  \sqrt{1 - \bar\alpha_t}\, \bm{\epsilon}$
  \State With \texttt{no\_grad}: compute aggregated
  features $\mbf{f}$ and $\mu_\theta \gets$ mean
  head$(\mbf{f})$
  \State $\Sigma_\phi \gets h_\phi(\mbf{f})$
  \Comment{diagonal in spatial or DCT basis}
  \State $\mbf{r} \gets \mbf{w} - \mu_\theta$
  \Comment{stationary target since $\mu_\theta$ frozen}
  \If{DCT variant}
    \State $\hat{\mbf{r}} \gets \mathrm{DCT}(\mbf{r})$
    \Comment{transform residual to DCT domain}
    \State $\ell_\phi^{(2)} \gets
    \tfrac12 \hat{\mbf{r}}^\top \Sigma_\phi^{-1} \hat{\mbf{r}}
    + \tfrac12 \log |\Sigma_\phi|$
  \Else
    \State $\ell_\phi^{(2)} \gets
    \tfrac12 \mbf{r}^\top \Sigma_\phi^{-1} \mbf{r}
    + \tfrac12 \log |\Sigma_\phi|$
  \EndIf
  \Comment{Equation~\eqref{eq:nll}}
  \State $\mc{L} \gets \ell_\phi^{(2)} +
  \lambda_{\mathrm{var}} \, \| \log \Sigma_\phi \|_2^2$
  \State Backpropagate $\mc{L}$ through $h_\phi$ only;
  clip $\| \nabla_\phi \mc{L} \|_2$ to $g_{\max}$;
  optimizer step on $\phi$.
\EndFor
\end{algorithmic}
\end{algorithm}

\begin{algorithm}[h]
\caption{\ours{} inference: covariance-aware native-latent guidance}
\label{alg:inference}
\begin{algorithmic}[1]
\Require Encoded measurement
$\mbf{w} = \mc{E}(\mby)$; noise level $\sigma_y$;
trained mean head $\mu_\theta$ and variance head
$h_\phi$; pretrained LDM denoiser; sampler with
schedule $\{\bar\alpha_t\}_{t=1}^{T}$; step size
$\eta$; per-coordinate gradient clip $c$.
\State Sample $\mbz_T \sim \mc{N}(\mbf{0}, \mbf{I})$.
\For{$t = T, T\!-\!1, \ldots, 1$}
  \State Run one unconditional sampler step from
  $\mbz_t$ to obtain $\mbz_{t-1}'$ and the Tweedie
  estimate $\mbz_{0 \mid t}$.
  \State Compute features $\mbf{f}$ at
  $(\mbz_t, t, \sigma_y)$ via the UNet;\quad obtain
  $\mu_\theta$ from the mean head and $\Sigma_\theta = \diag(v)$
  from $h_\phi(\mbf{f})$.
  \State $v \gets
  \mathrm{stop\text{-}grad}(v)$
  \Comment{detach covariance for guidance}
  \State $\mbf{r} \gets \mbf{w} - \mu_\theta$
  \If{DCT variant}
    \State $\hat{\mbf{r}} \gets \mathrm{DCT}(\mbf{r})$
    \State $\mbf{u} \gets \mathrm{DCT}^{-1}\!\bigl(
    \hat{\mbf{r}} \,/\, v\bigr)$
    \Comment{per-frequency precision weighting}
    \State $\mc{J}_t \gets \tfrac{1}{\mathrm{stop\text{-}grad}(\| \mbf{r} \|_2)}
    \langle \mbf{u}, \mbf{r} \rangle$
    \Comment{detach denominator}
  \Else
    \State $\mc{J}_t \gets \tfrac{1}{\mathrm{stop\text{-}grad}(\| \mbf{r} \|_2)}
    \bigl\|\,
    \mbf{r} \,/\, \sqrt{v} \,\bigr\|_2^2$
    \Comment{normalize by $\|\mbf{r}\|_2$}
  \EndIf
  \State $\mbf{g}_t \gets \nabla_{\mbz_t} \mc{J}_t$
  \Comment{gradient flows through $\mu_\theta$ only}
  \State $\mbz_{t-1} \gets \mbz_{t-1}' - \eta \cdot
  \mathrm{clip}(\mbf{g}_t, -c, c)$
\EndFor
\State \Return $\mc{D}(\mbz_0)$.
\end{algorithmic}
\end{algorithm}

% ============================================================
\FloatBarrier
\section{Extended Experiments}
\label{app:full_results}
% ============================================================

\FloatBarrier

\FloatBarrier
\subsection{Extended Qualitative Results}
\label{app:qual}

The following figures demonstrate extended qualitative results of our method on the FFHQ dataset. We observe that \ours{} consistently recovers crucial high-frequency detail that SILO misses, highlighting the importance of modeling uncertainty in guidance. STSL on average also retains some of these high-frequency details at the expense of significantly increased time. We observe that STSL often introduces certain artifacts, which may be due to the Jacobian of the decoder, as pointed out in \citet{raphaeli2025silo}. Note that the center box inpainting task we have on FFHQ is particularly challenging as the entire face is often occluded. This stress tests the need for powerful pretrained priors coupled with accurate guidance algorithms. We observe that baselines often do not preserve simple facial structure such has having two eyes, a nose, and a mouth. \ours{} usually preserves this, but sometimes struggles with smooth boundaries between the masked and unmasked region. On the JPEG decompression task, we clearly see the benefit of modeling covariance in the DCT domain, which gives high-quality reconstructions and drastically improves upon the spatial variant of \ours.

\begin{figure}[p]
    \centering
    \includegraphics[width=\linewidth]{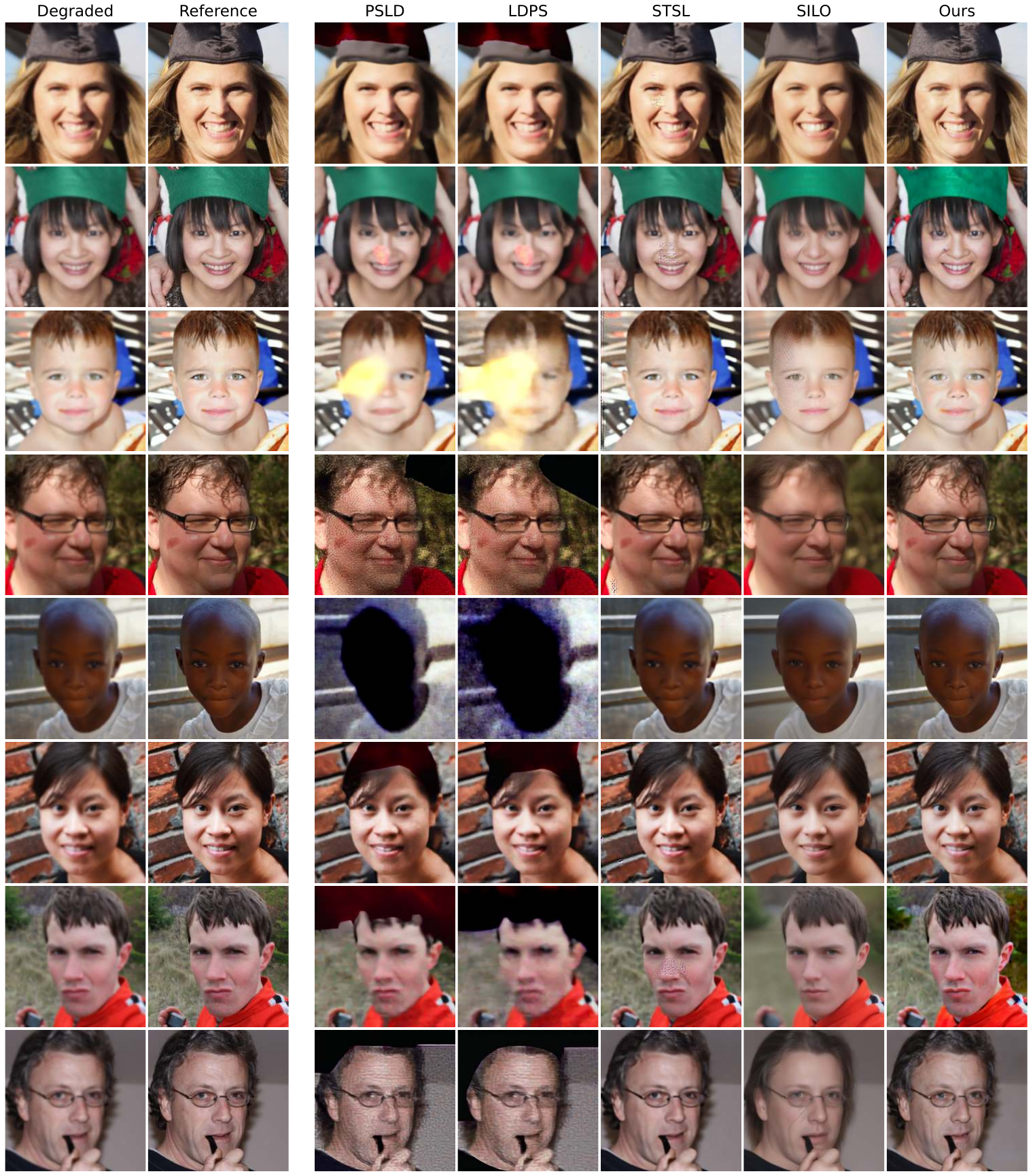}
    \caption{Gaussian Blur task on FFHQ dataset.}
    \label{fig:qual_gb}
\end{figure}

\begin{figure}[p]
    \centering
    \includegraphics[width=\linewidth]{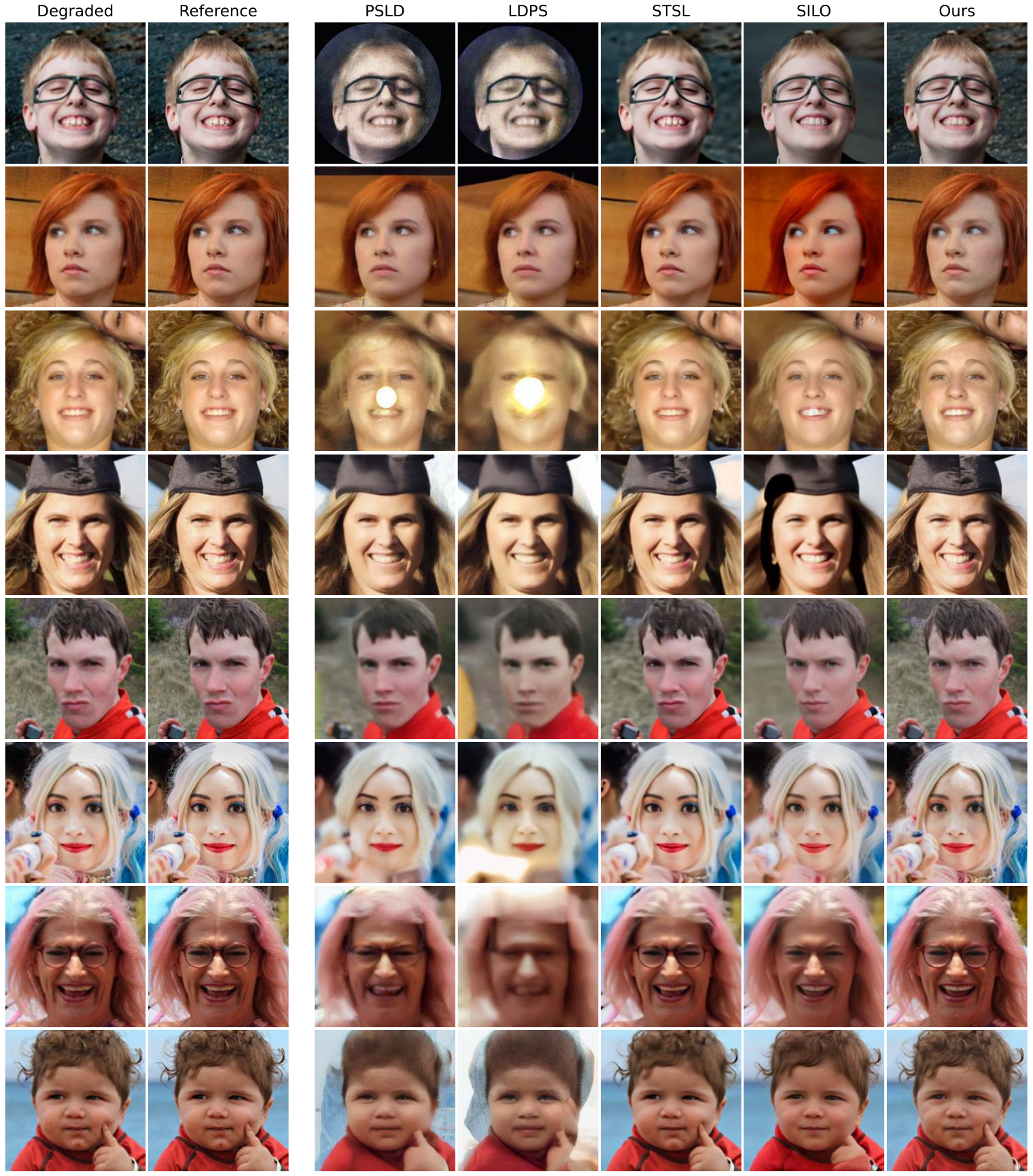}
    \caption{4x Super-Resolution task on FFHQ dataset.}
    \label{fig:qual_sr4}
\end{figure}

\begin{figure}[p]
    \centering
    \includegraphics[width=\linewidth]{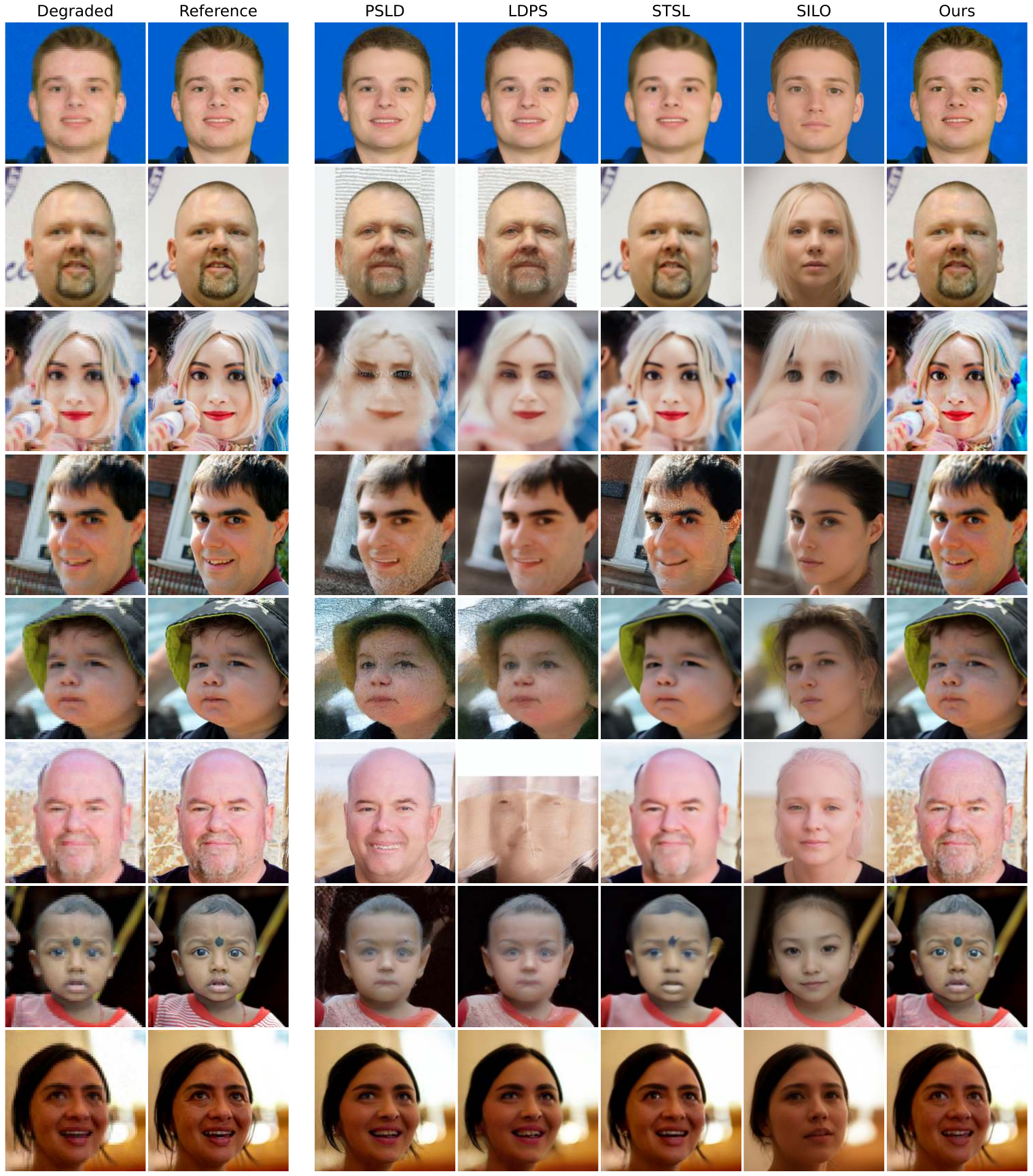}
    \caption{8x Super-Resolution task on FFHQ dataset.}
    \label{fig:qual_sr8}
\end{figure}

\begin{figure}[p]
    \centering
    \includegraphics[width=\linewidth]{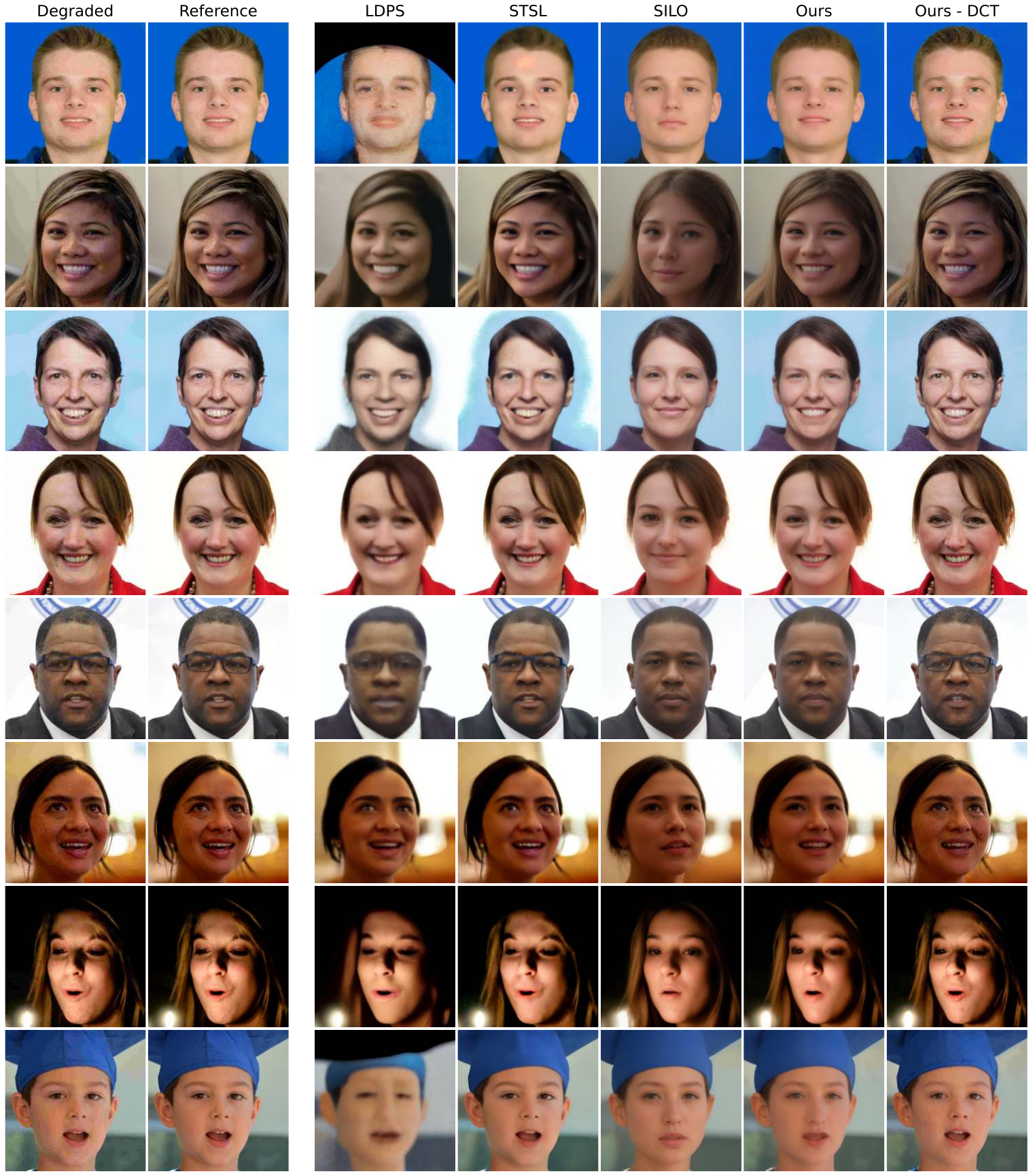}
    \caption{JPEG Decompression task on FFHQ dataset.}
    \label{fig:qual_jpeg}
\end{figure}

\begin{figure}[p]
    \centering
    \includegraphics[width=\linewidth]{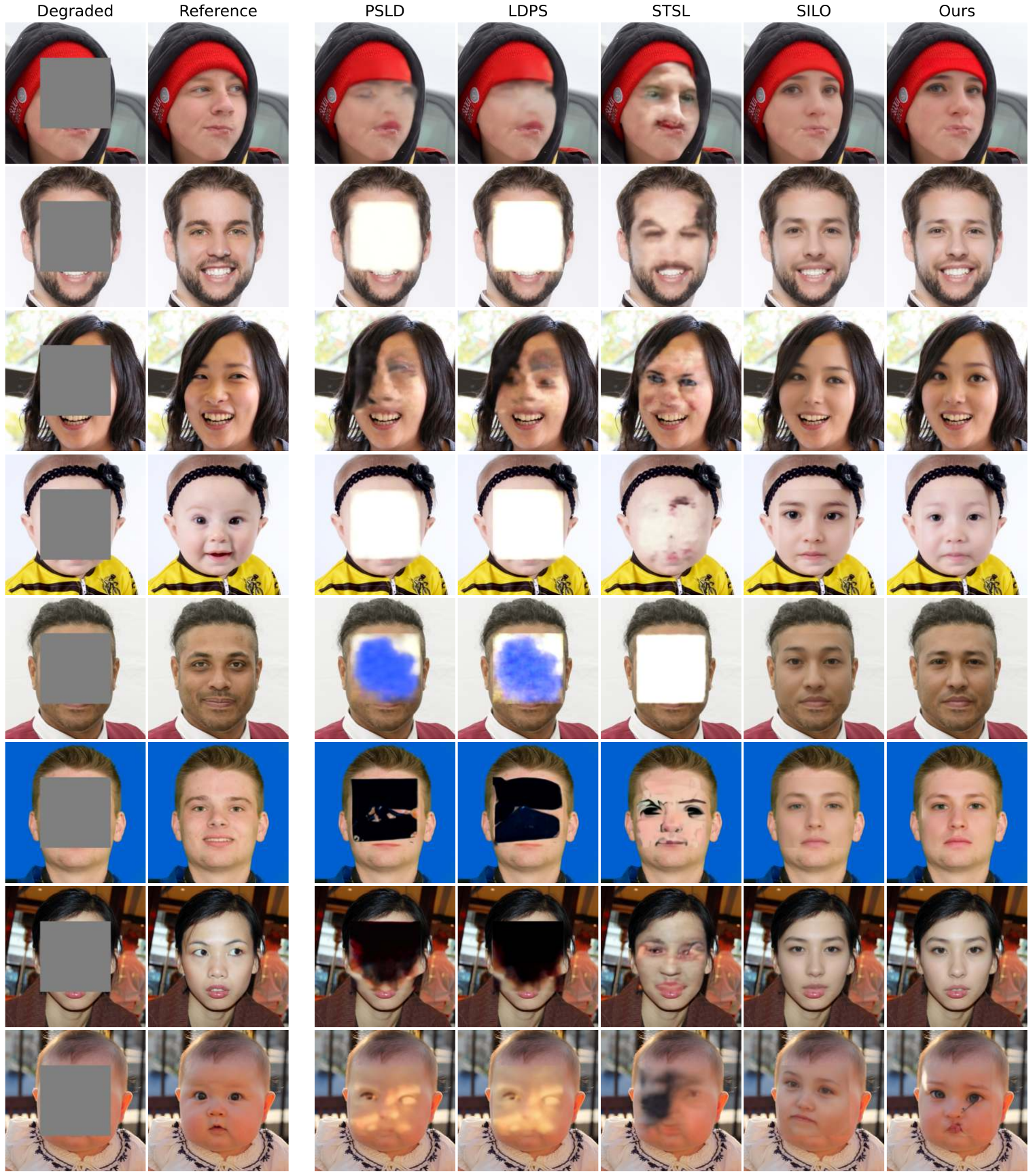}
    \caption{Box Inpainting task on FFHQ dataset.}
    \label{fig:qual_ip}
\end{figure} 

\FloatBarrier
\subsection{Extending to Richer Covariance Parameterizations}
\label{app:block_diag}

Our probabilistic framework in principle permits arbitrary covariance models. In the main text, spatial- and
DCT-diagonal covariances were chosen to show even simple covariance models can substantially improve upon baselines. Nonetheless, to
test a richer parameterization, in this section, we learn a dense $4\times4$ covariance across
latent channels at each DCT frequency $(u,v)$. We parameterize each covariance
block through its Cholesky factor to guarantee positive definiteness.

As shown in Table~\ref{tab:block_dct}, the block-DCT covariance can further improve over the DCT-diagonal model. These results demonstrate that explicitly modeling cross-channel correlations provides
further gains without sacrificing the efficiency advantage of the
native-latent formulation. We leave still richer covariance parameterizations
to future work.

\begin{table}[h]
    \centering
    \caption{
        \textbf{Richer covariance parameterizations improve restoration quality.}
    }
    \label{tab:block_dct}
    \small
    \begin{tabular}{llccc}
        \toprule
        Task & Method & FID $\downarrow$ & LPIPS $\downarrow$ & PSNR $\uparrow$ \\
        \midrule

        \multirow{4}{*}{Gaussian deblurring}
        & SILO                & 47.3 & 0.367 & 25.03 \\
        & \ours               & 29.8 & 0.326 & \textbf{27.34} \\
        & \ours-DCT           & 30.9 & 0.338 & 26.36 \\
        & \ours-DCT-block     & \textbf{22.0} & \textbf{0.292} & 26.48 \\
        \midrule

        \multirow{4}{*}{JPEG decompression}
        & SILO                & 68.3 & 0.471 & 20.87  \\
        & \ours               & 46.3 & 0.393 & 23.91  \\
        & \ours-DCT           & 27.3 & 0.326 & 26.26  \\
        & \ours-DCT-block     & \textbf{24.1} & \textbf{0.303} & \textbf{27.14} \\
        \bottomrule
    \end{tabular}
\end{table}

\FloatBarrier
\subsection{Probing Calibration of Uncertainty Head}
\label{app:calibration}

This section examines whether the learned diagonal covariance
$\Sigma_\theta$ obtained by minimizing the negative log-likelihood
in Eq.~\ref{eq:nll} produces well-calibrated uncertainty estimates on
held-out data, both on average and at the level of individual latent
coordinates.

\paragraph{Testing Setup.}
For each held-out measurement $\mby$ we compute the standardized residual
\begin{equation}\label{eq:zdef}
z_i \;=\; \frac{\mc{E}(\mby)_i \,-\, \mu_\theta(\mbz_t,\,t,\,\sigma_y)_i}
                {\sqrt{\Sigma_\theta(\mbz_t,\,t,\,\sigma_y)_{ii}}},
\end{equation}
indexed by latent coordinate $i \in \{1,\ldots,4\!\times\!64\!\times\!64\}$.
Under a calibrated diagonal Gaussian likelihood, $z_i \sim \mc{N}(0,1)$
at every coordinate, so $\mathbb{E}[z_i^2] = 1$ and the empirical variance
of $z_i$ across an i.i.d.\ batch of measurements is also $1$. We therefore
report two complementary quantities. First, the \emph{aggregate}
$\mathbb{E}[z^2]$ pooled over all coordinates---a necessary condition for
calibration. Second, the \emph{per-coordinate} empirical variance
$\mathrm{Var}_b[z_i]$ taken across the held-out batch and studied as a
distribution over $i$. The per-coordinate quantity probes calibration at
the level of individual latent coordinates; uniform calibration requires
this distribution to be concentrated at $1$ for every $i$.

For each task we evaluate the trained \ours{} checkpoint on $N\!=\!256$
held-out FFHQ images. For every test image we
apply the task-specific operator, add measurement noise with $\sigma_y$
drawn from $\mathcal{U}(0, 0.1)$, encode both the clean image and the
noisy measurement through the latent autoencoder, and condition the
covariance head on $(\mbz_t,\,t,\,\sigma_y)$. We then compute $z$ from
Eq.~\ref{eq:zdef} on the resulting tensor of shape $(N,\,4,\,64,\,64)$.
With $N\!=\!256$ samples, the sampling-noise floor on a per-coordinate
variance estimate around the true value of $1$ is
$\sqrt{2/(N-1)}\!\approx\!0.089$, which sets an irreducible lower bound
on the cross-coordinate spread of $\mathrm{Var}_b[z_i]$ that any
calibration test at this $N$ can resolve.

\paragraph{Calibration holds on all five tasks.}
Table~\ref{tab:calibration} summarises the results.
The aggregate $\mathbb{E}[z^2]$ lands within a few percent of $1$ on
every task. The per-coordinate empirical variance
$\mathrm{Var}_b[z_i]$ is also tightly distributed around $1$: the
cross-coordinate \emph{mean} ranges from $0.94$ to $1.03$, the
\emph{median} from $0.95$ to $1.01$, and on every task at least $99.5\%$
of the $16{,}384$ latent coordinates fall within a factor of two of
perfect calibration. Under the stricter window
$\mathrm{Var}_b[z_i]\!\in\![0.8,\,1.25]$ (within $\pm 25\%$ of the
calibrated value), $87\%$--$97\%$ of coordinates qualify. Because the
sampling-noise floor at $N\!=\!256$ is already $\approx\!0.09$, the
observed cross-coordinate spread of $0.10$--$0.20$ leaves only modest
room for systematic miscalibration.

\begin{table}[t]
\centering
\caption{Per-coordinate calibration of the learned diagonal covariance
$\Sigma_\theta$ on FFHQ ($N\!=\!256$ held-out images per task).
$\mathbb{E}[z^2]$, $\overline{z}$ and $\mathrm{std}(z)$ are pooled over
all latent coordinates. The remaining columns characterise the
distribution of $\mathrm{Var}_b[z_i]$ over the $4\!\times\!64\!\times\!64
= 16{,}384$ latent coordinates, and the per-coordinate residual bias
$|\overline{z_i}|$ averaged over coordinates. The sampling-noise floor on
$\mathrm{Var}_b[z_i]$ at $N\!=\!256$ is $\approx 0.089$.}
\label{tab:calibration}
\small
\setlength{\tabcolsep}{4pt}
\begin{tabular}{lcccccccc}
\toprule
Task & $\mathbb{E}[z^2]$ & $\overline{z}$ & $\mathrm{std}(z)$
     & \multicolumn{3}{c}{$\mathrm{Var}_b[z_i]$ across coords}
     & \multicolumn{1}{c}{$\Pr[\,\mathrm{Var}_b[z_i]\!\in\![0.8,1.25]\,]$}
     & $|\overline{z_i}|$ \\
\cmidrule(lr){5-7}
     &      &       &       & mean & median & std & & avg over coords \\
\midrule
\texttt{gb}   & 1.033 & $+0.024$ & 1.016 & 1.030 & 1.010 & 0.199 & 93.7\% & 0.068 \\
\texttt{sr4}  & 1.069 & $-0.139$ & 1.024 & 0.943 & 0.947 & 0.125 & 87.4\% & 0.326 \\
\texttt{sr8}  & 1.015 & $-0.074$ & 1.005 & 0.958 & 0.964 & 0.109 & 93.1\% & 0.213 \\
\texttt{ip}   & 0.993 & $+0.020$ & 0.996 & 0.992 & 0.981 & 0.137 & 93.6\% & 0.058 \\
\texttt{jpeg} & 0.999 & $-0.030$ & 0.999 & 0.997 & 0.992 & 0.102 & 97.0\% & 0.059 \\
\bottomrule
\end{tabular}
\end{table}

\begin{figure}[t]
\centering
\includegraphics[width=\linewidth]{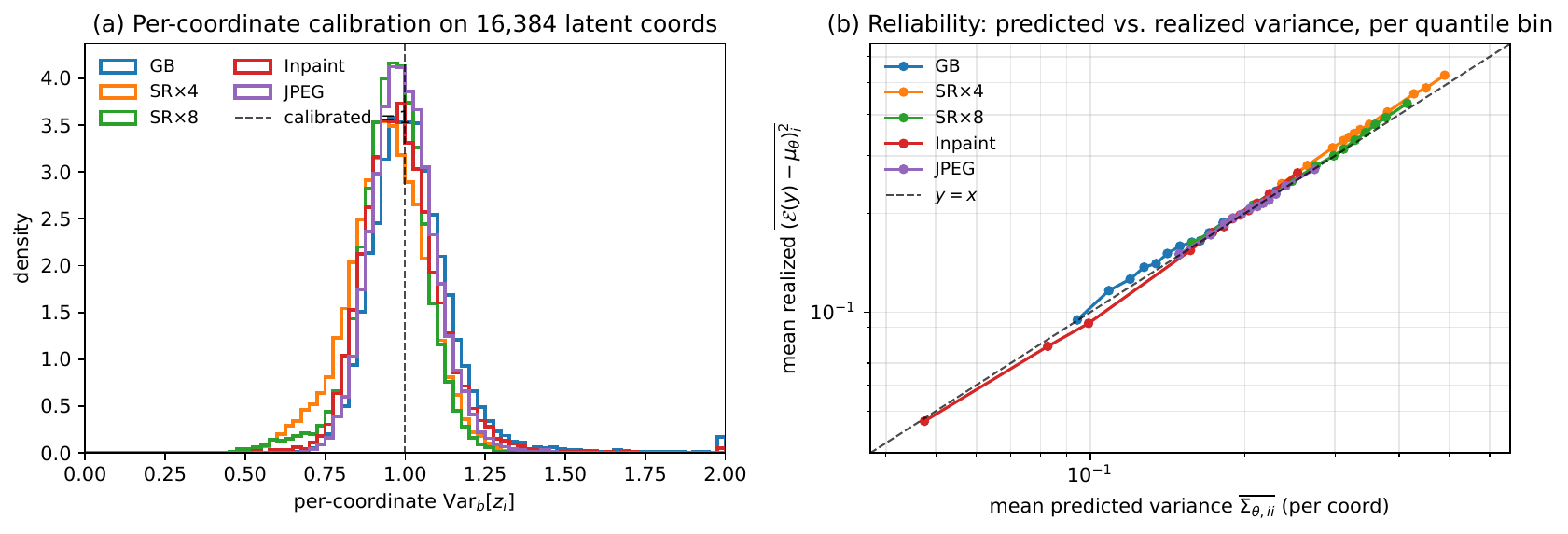}
\caption{Calibration of the learned diagonal covariance $\Sigma_\theta$
on FFHQ, all five tasks overlaid (\texttt{gb}, \texttt{sr4}, \texttt{sr8},
\texttt{ip}, \texttt{jpeg}; $N\!=\!256$ held-out images each).
\textbf{(a)} Distribution of the per-coordinate empirical variance
$\mathrm{Var}_b[z_i]$ over the $16{,}384$ latent coordinates; the dashed
line marks the calibrated value $1$. All five tasks are sharply peaked
at $1$. \textbf{(b)} Reliability diagram on log-log axes: coordinates
binned into 12 quantiles by mean predicted variance; the dashed line is
$y\!=\!x$. All five tasks track $y\!=\!x$ across roughly two orders of
magnitude of predicted variance, indicating that the variance head is
calibrated not only on average but conditional on its own predictions.}
\label{fig:calibration_combined}
\end{figure}

Figure~\ref{fig:calibration_combined}(a) overlays the distribution of
$\mathrm{Var}_b[z_i]$ over the $16{,}384$ latent coordinates for all five
tasks. Every distribution is sharply peaked at the calibrated value
$\mathrm{Var}_b[z_i]\!=\!1$ with no heavy tails on either side; the
visible width is dominated by the $N\!=\!256$ sampling-noise floor.
Figure~\ref{fig:calibration_combined}(b) is a reliability diagram:
latent coordinates are binned by predicted variance quantile, and the
mean predicted variance in each bin is plotted against the mean realised
squared residual on log-log axes. All five tasks track the
perfect-calibration line $y\!=\!x$ across roughly two orders of magnitude
of predicted variance, indicating that the variance head is calibrated
not only on average but also \emph{conditionally on its own
predictions}: the coordinates the model says it is uncertain about are
exactly the coordinates whose realised residuals are large, and
vice versa.

The cross-coordinate concentration of $\mathrm{Var}_b[z_i]$ around $1$
together with the alignment of the reliability diagram with $y\!=\!x$
provide empirical evidence that the learned diagonal covariance
$\Sigma_\theta$ is calibrated on held-out data, supporting the use of
$\Sigma_\theta^{-1}$ as a precision matrix in the guidance term of
Eq.~\ref{eq:guidance}.

\FloatBarrier
\subsection{Further Ablation Studies}
\label{app:further_ablations}

\begin{figure}[ht]
    \centering
    \includegraphics[width=\linewidth]
    {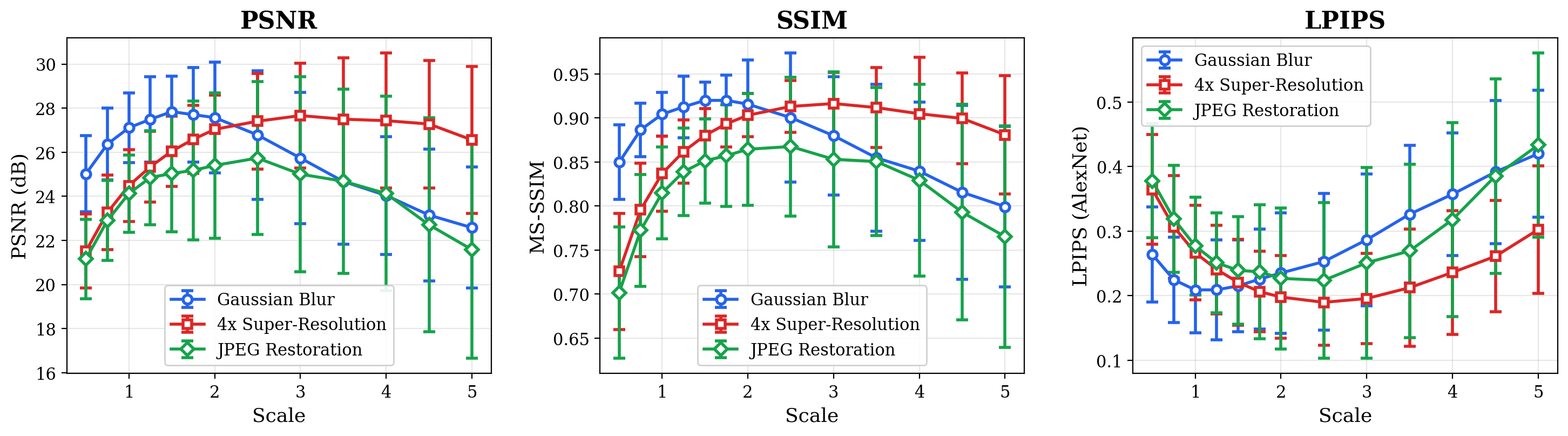}
    \caption{Sensitivity to the guidance scale
    hyperparameter on FFHQ dataset.}
    \label{fig:scale_sensitivity}
\end{figure}

\paragraph{Guidance scale sensitivity.}
On a held-out validation set of 100 images, Figure~\ref{fig:scale_sensitivity} shows the
sensitivity of \ours{} to the guidance
scale hyperparameter on FFHQ. We see that there exists a broad range of scales for which all metrics remain high. Our results were optimized for perceptual quality (LPIPS). 

\paragraph{Stop-gradient approximation.}

To evaluate the effect of stopping gradients through the learned covariance on
reconstruction quality, we compare our implementation against guidance using
the full gradient of the learned likelihood on 100 FFHQ images on the Gaussian deblurring task.
Results are shown in Table~\ref{tab:stopgrad}.

\begin{table}[h]
    \centering
    \caption{
        \textbf{Effect of stopping gradients through the learned covariance.}
    }
    \label{tab:stopgrad}
    \begin{tabular}{lccc}
        \toprule
        Guidance & FID $\downarrow$ & LPIPS $\downarrow$ & PSNR $\uparrow$ \\
        \midrule
        Stop-gradient (ours)       & 61.7          & \textbf{0.317} & \textbf{27.47} \\
        Full covariance gradient   & \textbf{60.9} & 0.325          & 27.44 \\
        \bottomrule
    \end{tabular}
\end{table}

Stop-gradient preserves essentially the same reconstruction quality while
providing substantially greater optimization stability. With the full
covariance gradient, we often observe large gradient-norm spikes that
destabilize guidance, resulting in a much narrower range of usable step sizes.
In contrast, stop-gradient remains stable across a broad range of guidance
scales, as shown in Figure \ref{fig:scale_sensitivity}.

\FloatBarrier
\section{Limitations and Broader Impact}
\label{app:limitations}

We discuss below some limitations of our work. 

The first set of limitations are of the broader framework of native-latent guidance. \ours{} inherits the limitation of SILO in that sufficiency can only hold when the measurement itself lies in pixel-space; for example, for tasks like phase retrieval, encoding the measurement via a pretrained image-space LDM will not work and clearly in this case, $\mc{E}(\mby)$ is not a sufficient statistic for the latent posterior. We hope that our characterization of sufficiency will inspire future work on learning the proper measurement transformation. Secondly, native-latent guidance trades off inference time with task-specific training. Our method does not directly address these limitations, but considers an orthogonal improvement: formulating a principled probabilistic framework for native-latent guidance and uncovering the role of covariance structure.

In practice, to maintain efficiency and for computational tractability, we restrict the learned covariance to be diagonal in spatial or DCT domain. We hope that future work may consider richer covariance models, such as low-rank approximations. 

On the theoretical front, we do not claim that our theoretical setting is perfectly indicative of real-world data. Rather, we aim to use the theoretical results as a motivating example for studying covariance structure in native-latent guidance. We point out an interesting connection between our autoencoder latent structure as a low-pass filter and recent work that studies the frequency properties of nonlinear LDMs \citep{skorokhodov2025improving}. It is an interesting direction to theoretically formalize the frequency structure of LDM latent spaces in nonlinear architectures.

%%%%%%%%%%%%%%%%%%%%%%%%%%%%%%%%%%%%%%%%%%%%%%%%%%%%%%%%%%%%

\end{document}